\documentclass[preprint,noshowkeys,amsmath,amssymb,superscriptaddress,nofootinbib, longbibliography]{revtex4-1}

\usepackage{bm}
\usepackage{slashed}
\usepackage{ulem}

\usepackage[pdftex]{graphicx,color}
\usepackage{epstopdf}
\usepackage[bookmarks=true,bookmarksnumbered=true,bookmarkstype=toc, colorlinks=true, citecolor=blue, linkcolor=blue]{hyperref} 
\usepackage{units}

\usepackage{mathtools}% http://ctan.org/pkg/mathtools
\usepackage{color}

\definecolor{green}{cmyk}{1,0,1,0}

\definecolor{pink}{cmyk}{0,0.5,0,0}

\definecolor{pastelpink}{cmyk}{0,0.25,0,0}

\definecolor{softpink}{cmyk}{0,0.125,0,0}

\definecolor{purple}{cmyk}{0.5,1.0,0.1,0}

\definecolor{violet}{cmyk}{0.75,1,0.25,0}

\newcommand{\nn}{\nonumber}

\usepackage[utf8]{inputenc}

\usepackage[geometry]{ifsym}

\begin{document}
%%% preprint numbers
\preprint{UME-PP-028}
\preprint{KYUSHU-HET-288}

%%%%% declaration for front matter%%%%%%%%%%%%%%%%%%%%%%%%%%%%%%%%%%%
\title{Dark photon pair production via off-shell dark Higgs at FASER}

\author{Takeshi Araki}
\email{t-araki@den.ohu-u.ac.jp}
\affiliation{Faculty of Dentistry, Ohu University, 31-1 Misumidou, Tomita-machi, Koriyama, Fukushima 963-8611, Japan}

\author{Kento Asai}
\email{kento@icrr.u-tokyo.ac.jp}
\affiliation{Institute for Cosmic Ray Research (ICRR), The University of Tokyo, Kashiwa, Chiba 277--8582, Japan}

\author{Yohei Nakashima}
\email{nakashima.youhei.775@s.kyushu-u.ac.jp}
\affiliation{Department of Physics, Kyushu University,
744 Motooka, Nishi-ku, Fukuoka, 819-0395, Japan}

\author{Takashi Shimomura}
\email{shimomura@cc.miyazaki-u.ac.jp}
\affiliation{Faculty of Education, Miyazaki University, Miyazaki, 889-2192, Japan}

%%%%% typeset front matter (including abstract) %%%%%%%%%%%%%%%%%%%%%%
\begin{abstract}
We consider a dark photon model in which the dark U(1) gauge symmetry is spontaneously broken by a vacuum expectation value of a new scalar boson.
We focus on the ForwArd Search ExpeRiment (FASER) and calculate its sensitivity to the dark photon produced from the off-shell decay of the new scalar boson.
It is found that the off-shell production extends the sensitivity region beyond the kinematical threshold of the on-shell decay of the scalar boson, and that the sensitivity region can be spanned to unexplored region.
We also show the parameter space in which perturbative calculation is valid for the unitarity of an S matrix.
\end{abstract}

\date{\today}

\maketitle

%%%%%%%%%%%%%%%%%%%%%%%%%
\section{Introduction}
%%%%%%%%%%%%%%%%%%%%%%%%%
Feebly interacting and light dark sector has received growing interests in recent years. 
It is motivated by unanswered problems in the Standard Model (SM) such as dark matter, neutrino masses, 
strong CP problem, and so on (for review see ref.~\cite{Antel:2023hkf}). 
Ordinary matters interact with the dark sector only via so-called portal particles whose masses are postulated below GeV scale.
Such light portal particles are searched in present and future high intensity experiments. 
Since many extensions beyond the minimal setup have been proposed, 
detailed studies on the production of portal particles are important to determine the sensitivity to the dark sector in experiments.

There are generally four types of portal particles; vector, scalar, fermion, and axion-like particles. 
Among them, vector portal refers to as gauge bosons of additional gauge symmetry \cite{Holdom:1985ag, Foot:1990mn, He:1990pn, He:1991qd, Kopp:2012dz, Heeck:2014zfa, Bilmis:2015lja, Jeong:2015bbi} (for review see refs.~\cite{Ilten:2018crw,Bauer:2018onh}). The simplest one is 
the vector boson of an Abelian gauge symmetry U(1)$_D$ under which all the SM particles are singlet. Such a vector portal can mix kinetically  with the hypercharge gauge boson 
\cite{Okun:1982xi,Galison:1983pa,Holdom:1985ag,Foot:1991kb,Babu:1997st}.
Due to the kinetic mixing, it can interact with the SM particles through the electromagnetic current. 
In this sense, this vector portal is called dark photon \cite{Pospelov:2007mp,Huh:2007zw,Pospelov:2008zw}. 
The dark photon can be directly produced from bremsstrahlung, meson decays, and QCD processes through the electromagnetic current in terrestrial experiments.
On the other hand, in general the dark photon has mass. One of the origins of the dark photon mass can be spontaneous symmetry breaking of U(1)$_D$. For this mechanism to occur, complex scalar bosons or dark Higgs bosons are introduced to break the symmetry spontaneously. Then, the dark photon mass is generated by the vacuum expectation value of the dark Higgs field.
As a direct consequence of the mass generation, the dark Higgs boson can decay into a pair of dark photons. 
This decay can dominate the dark Higgs boson decays due to the enhancement by longitudinal polarization of massive dark photon.
In such a situation, the dark Higgs decay becomes new production process of dark photon.
In refs.~\cite{Araki:2020wkq, Araki:2022xqp, Foguel:2022unm, Cheung:2024oxh}, the dark photon production from the dark Higgs decay 
was studied. 
Using a large number of the dark Higgs bosons produced by heavy meson decays through the scalar mixing with the SM Higgs boson, 
it was shown that the sensitivity to the dark photon parameters can be enlarged. 
The similar analyses have been studied for inelastic dark matter \cite{Li:2021rzt}, lepton flavor violation \cite{Araki:2022xqp},  
dark Higgs \cite{Foguel:2022unm, Ferber:2023iso, Felkl:2023nan} in gauged $L_\mu - L_\tau$ \cite{Nomura:2020vnk} and $B-L$ models \cite{Dev:2021qjj}. 
In these studies, only decay of the on-shell dark Higgs was considered.

When the dark Higgs boson is heavier than the mesons, it cannot be produced from the meson decays. 
Even if the dark Higgs boson is lighter than the meson, it cannot decay into the dark photons  when the dark photon mass is two times heavier than the dark Higgs boson mass. 
In this situation, however, the mesons can decay directly into the dark photons through off-shell state of the dark Higgs boson. 
Such meson decays can be substantial due to the enhancement by the longitudinal polarization. 
In fact, suppression by the propagator of the dark Higgs boson can be compensated by the contribution from longitudinal polarization\footnote{Similar case was studied for a pair production of the dark Higgs boson through the off-shell SM Higgs boson in meson decay \cite{Feng:2017vli}. }.
On the other hand, dark matter contained in the dark sector can interact with SM particles through the dark photon. 
When the dark matter is produced from freeze-out mechanism in the early Universe, the gauge coupling constant of U(1)$_D$ must 
be of order unity to satisfy the observed dark matter abundance \cite{Izaguirre:2013uxa, Batell:2014mga, Kahn:2014sra, Izaguirre:2015yja}.
Such a large gauge coupling will also result in substantial number of dark photons via off-shell state of the dark Higgs boson.
Therefore, the sensitivity to the dark photon parameters should be studied in the case of the off-shell dark Higgs boson. 
In addition, as we will show, the perturbative unitarity bound \cite{Lee:1977yc, Lee:1977eg} suggests similar order of the mass of the dark Higgs boson to that of the dark photon for the large gauge coupling.  
Thus, it will be natural to consider the light dark photon and dark Higgs boson simultaneously.

The dark photon and dark Higgs boson have been searched in collider and fixed target/beam dump experiments as well as astrophysical observations. 
To date, the parameter space of the dark photon and dark Higgs boson have been tightly constrained. For complete lists of the experiments and constraints, 
we refer the reader to 
review for instance \cite{Bauer:2018onh, Ilten:2018crw} and \cite{Winkler:2018qyg,Ferber:2023iso}. Further searches are conducted in experiments 
such as Belle-II \cite{Belle-II:2018jsg}, NA64 \cite{NA64:2019auh}, LHCb \cite{LHCb:2016awg, LHCb:2019vmc}. 
Among these, the FASER (ForwArd Search ExpeRiment) experiment 
\cite{Feng:2017uoz, Feng:2017vli, FASER:2018ceo, FASER:2018eoc, FASER:2018bac} exploits the high energy 
proton-proton collision at LHC to search for neutral long-lived particles. 
The detector is placed 480 m downstream from the ATLAS interaction point (IP). Utilizing a large cross-section of proton-proton 
inelastic interaction in the forward direction, FASER realizes high sensitivity to new light and long-lived particles. 
FASER and its associated detector FASER$\nu$~\cite{FASER:2019dxq,FASER:2020gpr} started data taking since 2022 and will collect about 150 fb$^{-1}$ of data with 14 TeV collision energy in LHC Run 3. 
Both experiments already reported their first result on dark photon search \cite{FASER:2023tle}, and high-energy neutrino interaction cross section \cite{FASER:2021mtu, FASER:2023zcr, FASER:2024hoe}, respectively.
To enlarge its sensitivity and reach, an upgrade to FASER2 with larger detector is planned for High-Luminosity LHC era, aiming to accumulate 3 ab$^{-1}$ \cite{Anchordoqui:2021ghd, Feng:2022inv}.

The purpose of this paper is to show that the dark photon production via off-shell dark Higgs boson gives a substantial number of events. 
The sensitivity region at FASER2 can be enlarged when this production is taken into account. It is also shown that the SM Higgs boson invisible decay and 
perturbative unitarity bounds can constrain the parameter space when the dark Higgs boson has sub-GeV mass. 

This paper is organized as follows. In section~\ref{sec:dark-photon-model}, we introduce the dark photon model with dark Higgs boson, and present 
the relevant interaction Lagrangian to our analyses. 
In section~\ref{sec:number-of-events}, we give the differential branching ratio of bottom quark into the dark photons 
via off-shell dark Higgs boson. Then, the formula to calculate expected number of events is shown. Constraints from the invisible decay of the SM Higgs boson and perturbative unitarity are given in section~\ref{sec:constraints}. In section~\ref{sec:sensitivity-region}, expected sensitivity plots to dark photon decay signal at FASER2 are presented, and we summarize our analysis in section~\ref{sec:conclusion}.
In appendix, we give the decay widths of the dark Higgs boson and dark photon, and also their lifetimes, respectively.

%%%%%%%%%%%%%%%%%%%%%%%%%%%%%%%%
\section{Dark Photon Model} \label{sec:dark-photon-model}
%%%%%%%%%%%%%%%%%%%%%%%%%%%%%%%%
We begin our discussion by deriving  the interaction Lagrangian used in our analysis. A dark photon model considered in this study 
is invariant under the gauge symmetries $G_{\mathrm{SM}} \times \mathrm{U(1)}_D$ where $G_{\mathrm{SM}}$ represents the SM gauge symmetry. 
The new gauge symmetry U(1)$_D$ is the symmetry of dark sector and all of the SM fields are assumed to be uncharged under this symmetry. 
Although the gauge boson $X_\mu$ of U(1)$_D$ has no direct interactions to the SM sector, the gauge symmetry allows $X_\mu$ to form 
the kinetic mixing term with the hypercharge gauge boson $B_\mu$ \cite{Okun:1982xi,Galison:1983pa,Holdom:1985ag,Foot:1991kb,Babu:1997st}. 
Then, the dark photon emerges as a linear combination of 
$X_\mu$ and $B_\mu$. In this study, the dark symmetry is assumed to be spontaneously broken by a new complex scalar boson $\Phi$. 
For the sake of minimality,  we consider the case that $\Phi$ is charged only under U(1)$_D$ and its gauge charge is assigned to 
be unity.

%%%%%%%%%%%%%%%%%%%%%%%%%
\subsection{Lagrangian}
%%%%%%%%%%%%%%%%%%%%%%%%%
The Lagrangian of the model is given by 
\begin{align}
\mathcal{L} = \mathcal{L}_{\mathrm{SM}} - \frac{1}{4} X^{\mu \nu} X_{\mu \nu}  + \frac{\epsilon}{2} B^{\mu\nu} X_{\mu\nu} 
+ |D_\mu \Phi|^2 - V(H,\Phi), \label{eq:full-lag}
\end{align}
where $\mathcal{L}_{\mathrm{SM}}$ denotes the SM Lagrangian without the Higgs potential. The second and third terms are the kinetic term 
of $X_\mu$ and kinetic mixing term between $X_\mu$ and $B_\mu$ with the constant parameter $\epsilon$. 
The field strength tensors of $X_\mu$ and $B_\mu$ are given by 
\begin{align}
X_{\mu\nu} = \partial_\mu X_\nu - \partial_\nu X_\mu,~~~B_{\mu\nu} = \partial_\mu B_\nu - \partial_\nu B_\mu,
\end{align}
respectively. The fourth term is the kinetic term of $\Phi$. The covariant derivative acting on $\Phi$ is given by
\begin{align}
D_\mu = \partial_\mu -i g' X_\mu,
\end{align}
 where $g'$ is the gauge coupling constant of U(1)$_D$.
The last term $V(H,\Phi)$ is the scalar potential in which $H$ is the electroweak SU(2) doublet field. 
The concrete form of $V$ is given by
\begin{align}
    V(H, \Phi) = -\mu_H^2 H^\dagger H -\mu_\Phi^2 \Phi^\dagger \Phi + \frac{\lambda_H}{2} (H^\dagger H)^2 
    + \frac{\lambda_\Phi}{2} (\Phi^\dagger \Phi)^2 + \lambda_{H\Phi} (H^\dagger H) (\Phi^\dagger \Phi), \label{eq:scalar-pot}
\end{align}
where $\mu^2_H$ and $\mu^2_\Phi$ as well as the quartic couplings $\lambda_H$, $\lambda_\Phi$ and $\lambda_{H\Phi}$ are taken to 
be positive. The values of these parameters are assumed to be chosen appropriately so that the electroweak and dark symmetry are 
broken spontaneously as desired.

%%%%%%%%%%%%%%%%%%%%%%%%%%%%%%%%
\subsection{Mass eigenstates of the scalar and gauge field}
%%%%%%%%%%%%%%%%%%%%%%%%%%%%%%%%
On the vacuum, the scalar fields $H$ and $\Phi$ are expanded around their vacuum expectation values $v$ and $v_\Phi$, respectively, as
\begin{align}
    H = \frac{1}{\sqrt{2}} 
    \begin{pmatrix}
        0 \\ v + H'
    \end{pmatrix}, ~~~~
    \Phi = \frac{1}{\sqrt{2}}( v_\Phi + \Phi'), \label{eq:scalar-expand}
\end{align}
where $H'$ and $\Phi'$ are the dynamical degrees of freedom. Here, the Nambu-Goldstone bosons to be absorbed by the gauge bosons are omitted.
Inserting eq.~\eqref{eq:scalar-expand} into the scalar potential eq.~\eqref{eq:scalar-pot}, we can obtain the scalar fields $h$ and $\phi$ 
in the mass basis 
\begin{align}
    \begin{pmatrix}
    H' \\ \Phi'
\end{pmatrix}
=\begin{pmatrix}
        \cos\alpha h + \sin\alpha \phi \\
        -\sin\alpha h + \cos\alpha \phi
    \end{pmatrix},
\label{eq:scalar-mass-eigenstates}
\end{align}
where the scalar mixing angle $\alpha$ is defined by
\begin{align}
    \tan 2 \alpha = -\frac{2 \lambda_{H\Phi} v v_\Phi}{\lambda_H v^2 - \lambda_{\Phi} v_\Phi^2}.
\end{align}
The masses of $h$ and $\phi$ are given by
\begin{subequations}
    \begin{align}
        m_h^2 &= \lambda_H v^2 c_\alpha^2 + \lambda_\Phi v_\Phi^2 s_\alpha^2 - 2 \lambda_{H\Phi} v v_\Phi s_\alpha c_\alpha, \\
        m_\phi^2 &=\lambda_H v^2 s_\alpha^2 + \lambda_\Phi v_\Phi^2 c_\alpha^2 + 2 \lambda_{H\Phi} v v_\Phi s_\alpha c_\alpha,
    \end{align}
\end{subequations}
where $s_\alpha$ and $c_\alpha$ are abbreviations for $\sin \alpha$ and $\cos \alpha$.
In our analysis, we take $\alpha$ to be small and identify $h$ as the SM Higgs boson with $m_h = 125$ GeV.

The mass eigenstates of the neutral gauge bosons $A_\mu,~Z_\mu$ and $A'_\mu$ are obtained by diagonalizing the mass term and kinetic term 
simultaneously, which are given by
\begin{align}
    \begin{pmatrix}
        \tilde{A}_\mu \\ \tilde{Z}_\mu \\ X_\mu
    \end{pmatrix}
    =
    \begin{pmatrix}
        1 & \epsilon c_W r s_\chi & \epsilon c_W r c_\chi \\
        0 & c_\chi - \epsilon s_W r s_\chi & -s_\chi - \epsilon s_W r c_\chi \\
        0 & r s_\chi & r c_\chi
    \end{pmatrix}
    \begin{pmatrix}
        A_\mu \\ Z_\mu \\ A'_\mu
    \end{pmatrix}, \label{eq:gauge-mass-eigenstates}
\end{align}
where $\tilde{A}_\mu$ and $\tilde{Z}_\mu$ denote the SM photon and neutral weak boson, respectively, in the limit of $\epsilon \to 0$. 
The gauge mixing angle is abbreviated as $s_\chi = \sin\chi$ and $c_\chi = \cos\chi$, and $r = (1-\epsilon^2)^{-1/2}$. 
Similarly, $s_W = \sin \theta_W$ and $c_W = \cos\theta_W$ for the Weinberg angle. 
The gauge mixing angle is defined by
\begin{align}
        \tan 2 \chi = \frac{2 \epsilon r s_W m_{\tilde{Z}}^2}{r^2 m_{X}^2 - (1-\epsilon^2 r^2 s_W^2)m_{\tilde{Z}}^2}, \label{eq:gauge-mixing}
\end{align}
where $m_{\tilde{Z}}$ is the SM $Z$ boson mass and 
\begin{align}
m_X = g' v_\Phi.
\end{align}
The masses of $Z$ and $A'$ are given by
\begin{subequations}
    \begin{align}
        m_{Z}^2 &= m_{\tilde{Z}}^2 c_\chi^2 + r^2( m_{X}^2 + \epsilon^2 s_W^2 m_{\tilde{Z}}^2) s_\chi^2 -2 \epsilon r s_W m_{\tilde{Z}}^2 s_\chi c_\chi, \\
        m_{A'}^2 &= m_{\tilde{Z}}^2 s_\chi^2 + r^2( m_{X}^2 + \epsilon^2 s_W^2 m_{\tilde{Z}}^2) c_\chi^2 +2 \epsilon r s_W m_{\tilde{Z}}^2 s_\chi c_\chi.
    \end{align}
\label{eq:gauge-mass-eigenvalues}
\end{subequations}

%%%%%%%%%%%%%%%%%%%%%%%%%
\subsection{Yukawa interaction of dark Higgs boson}
%%%%%%%%%%%%%%%%%%%%%%%%%
The dark Higgs boson can interact with the SM fermions through the scalar mixing. The interaction Lagrangian is obtained by 
inserting eq.\eqref{eq:scalar-mass-eigenstates} into the Yukawa interaction terms as 
\begin{align}
\mathcal{L}_{\mathrm{yukawa}} &= \sum_f \frac{m_f}{v} \overline{f} f (\cos\alpha h + \sin\alpha \phi), \label{eq:dark-higgs-yukawa}
\end{align}
where $f$ runs over all of the SM fermions and $m_f$ is the mass of the fermion $f$. Through this interaction, the dark Higgs boson can decay into the SM particles. The widths into the fermions and hadrons are explained in appendix~\ref{apdx:dark-higgs}. These decay widths are suppressed by $\sin^2\alpha$, and hence the dark Higgs boson can be long-lived when $\alpha$ is small. In figure~\ref{fig:DecayLengthOfDarkHiggs} in appendix~\ref{apdx:dark-higgs}, the decay length of the dark Higgs boson only with eq.~\eqref{eq:dark-higgs-yukawa} is shown for $\alpha = 10^{-4}$. The dark Higgs boson can travel over $1$ cm below $m_\phi = 1$ GeV in this case.

%%%%%%%%%%%%%%%%%%%%%%%%%%%%%%%%
\subsection{Gauge interaction of dark photon and dark Higgs boson}
%%%%%%%%%%%%%%%%%%%%%%%%%%%%%%%%
In the parameter space of our interest, the conditions that $\epsilon \ll 1$ and $m_{A'}^2 \ll m_{Z}^2$ are generally satisfied.  
Then, the gauge mixing angle eq.~\eqref{eq:gauge-mixing} is approximated as
\begin{align}
\chi \simeq  - \epsilon s_W + \mathcal{O}(\epsilon^3).
\end{align}
The mass eigenstates eq.~\eqref{eq:gauge-mass-eigenstates} become
\begin{align}
    \begin{pmatrix}
        \tilde{A}_\mu \\ \tilde{Z}_\mu \\ X_\mu
    \end{pmatrix}
    \simeq 
    \begin{pmatrix}
        1 & 0 & \epsilon c_W \\
        0 & 1 &  0 \\
        0 & - \epsilon s_W & 1
    \end{pmatrix}
    \begin{pmatrix}
        A_\mu \\ Z_\mu \\ A'_\mu
    \end{pmatrix},
    \label{eq:gauge-mass-eigenstates2}
\end{align}
and their masses eq.~\eqref{eq:gauge-mass-eigenvalues} become 
\begin{align}
        m_{Z}^2 \simeq m_{\tilde{Z}}^2,~~~m_{A'}^2 \simeq m_{X}^2,
\end{align}
up to the order of $\epsilon$, respectively.
Thus, the interaction term of dark photon with the SM fermions is simplify obtained by replacing $\tilde{A}_\mu \to A_\mu + \epsilon c_W A'_\mu$ 
in the SM Lagrangian as
\begin{align}
\mathcal{L}_{A'-\mathrm{fermion}} = \varepsilon e J^\mu_{\mathrm{em}} A'_\mu,
\label{eq:dp-Jem}
\end{align}
where $J^\mu_{\mathrm{em}}$ is the electromagnetic current for 
the fermions and $\varepsilon$ is defined by $\epsilon c_W$. 
With this interaction, dark photon can decay into the SM charged particles. The decay widths of the dark photon into the SM fermions and hadrons are given in appendix~\ref{apdx:dark-photon}. Figure~\ref{fig:DecayLengthOfDarkPhoton} shows the decay length of the dark photon normalized by $\varepsilon^2$. One can see from the figure that the dark photon becomes so long-lived that it can travel over several hundred meters for small kinetic mixing. Thus, such a dark photon can be a target of the FASER experiment.

The gauge interactions of the scalar bosons in mass basis are obtained from the scalar kinetic terms by using  
eqs.~\eqref{eq:scalar-mass-eigenstates} and \eqref{eq:gauge-mass-eigenstates2}. The relevant interactions in our analysis are 
given by
\begin{align}
\mathcal{L}_{\mathrm{gauge-scalar}} &= 
\frac{1}{2} {g'}^2 A'_\mu A'^\mu (-s_\alpha h + c_\alpha \phi)^2 
+ g' m_{A'}A'_\mu A'^\mu (-s_\alpha h + c_\alpha \phi).
\label{eq:lag-gauge-scalar}
\end{align}
The last term represents the decay of $\phi$ into a dark photon pair, which we 
analyze in this paper.
The decay width into two dark photons is given in appendix~\ref{apdx:dark-higgs}. The decay width can be enhanced due to the longitudinal polarization of the dark photons and dominate over the total decay width. The decay length of the dark Higgs boson is shown in figure~\ref{fig:DecayLengthOfDarkHiggs}. It can be seen that the decay length becomes suddenly short once the decay into dark photons is open. For the gauge coupling of order $0.1$, the decay length is so short that the dark Higgs boson promptly decay at the ATLAS IP.

%%%%%%%%%%%%%%%%%%%%%%%%
\subsection{Scalar self-interactions}
%%%%%%%%%%%%%%%%%%%%%%%%
In the end of this section, we give the scalar self-interaction Lagrangian which is derived from the scalar potential eq.~\eqref{eq:scalar-pot}. 
Inserting eqs.~\eqref{eq:scalar-expand} and \eqref{eq:scalar-mass-eigenstates} into eq.\eqref{eq:scalar-pot}, one obtains 
\begin{align}
\mathcal{L}_{\mathrm{scalar-self}}
    &= - \frac{\lambda_H}{8} \big[ 4 v (c_\alpha h + s_\alpha \phi)^3 + (c_\alpha h + s_\alpha \phi)^4 \big] \nn \\
    &\quad - \frac{\lambda_\Phi}{8} \big[ 4 v_\Phi (-s_\alpha h + c_\alpha \phi)^3 + (-s_\alpha h + c_\alpha \phi)^4 \big] \nn \\
    &\quad - \frac{\lambda_{H\Phi}}{4} \big[ 2 v (c_\alpha h + s_\alpha \phi) (-s_\alpha h + c_\alpha \phi)^2 + 2 v_\Phi (c_\alpha h + s_\alpha \phi)^2 (-s_\alpha h + c_\alpha \phi) \nn \\
    &\qquad \qquad+(c_\alpha h + s_\alpha \phi)^2 (-s_\alpha h + c_\alpha \phi)^2 \big].
\label{eq:lag-scalar}
\end{align}
The scalar self-interactions are used to derive the perturbative unitarity constraints.

%%%%%%%%%%%%%%%%%%%%%%%%%%%%%%%%%%%%%%%%%%%%%%%
\section{Expected Number of Dark Photon Decay at FASER} \label{sec:number-of-events}
%%%%%%%%%%%%%%%%%%%%%%%%%%%%%%%%%%%%%%%%%%%%%%%
The expected number of dark photon decays is given in this section. 
We study the situation where the dark photons are produced 
from the decays of the dark Higgs boson and consider the detection at the FASER and FASER2 experiment \cite{Feng:2017uoz,Feng:2017vli,Kling:2018wct,Feng:2018pew,FASER:2018ceo,FASER:2018eoc,FASER:2018bac,Feng:2022inv,FASER:2023tle}. 
If the dark Higgs boson is light enough, it can be produced as on-shell state from meson decays, such as $B$, $K$, $\eta$ and $\eta'$. 
Among these, the $B$ meson is dominant source for the dark Higgs boson.  
The $K_L$ meson is sub-dominant because it is absorbed by LHC neutral target absorbers before its decay \cite{Feng:2017vli}. 
The contributions from other mesons are almost negligible due to suppressed branching fraction.
For the production of the off-shell dark Higgs boson, the $B$ meson decay also will be dominant one because the decay vertex of the dark Higgs boson is 
common in all mesons as shown in figure~\ref{fig:B-meson-decay}.
Thus, we evaluate the $B$ meson decay in the following analyses.

%%%%%%%%%%%%%%%%%%%%%%%%%%%%%%%%
\subsection{Differential decay width of $B$ meson into Dark Photons}
%%%%%%%%%%%%%%%%%%%%%%%%%%%%%%%%
%%%%
\begin{figure}[t]
    \includegraphics[scale=0.3]{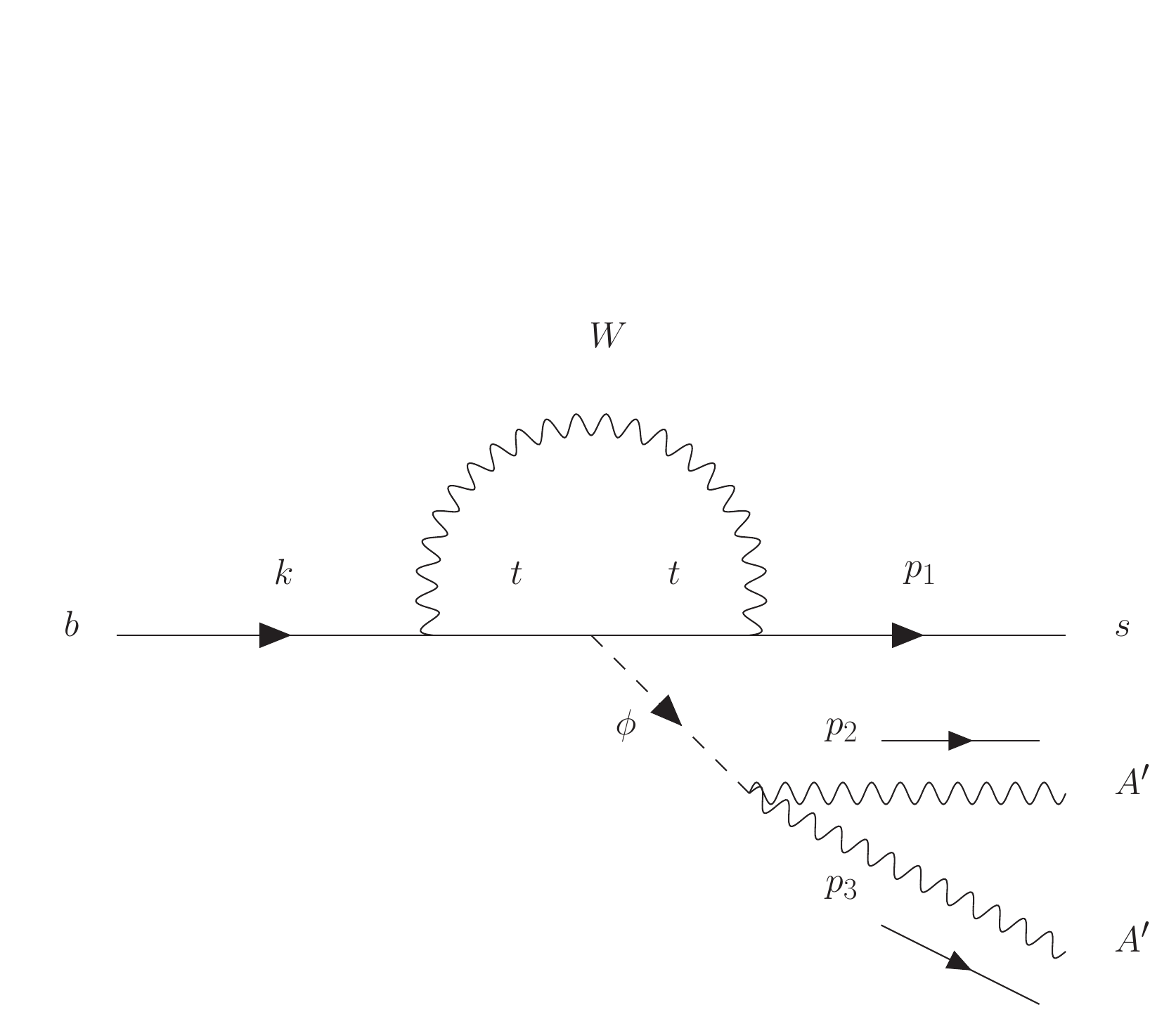}
    \caption{Feynman diagram of the dark photon pair production through the off-shell dark Higgs boson in $b$ quark decay.}
\label{fig:B-meson-decay}
\end{figure}
%%%%
The decay of $B$ meson into two dark photons through the off-shell dark Higgs boson can occur in loop-process shown in figure~\ref{fig:B-meson-decay}. 
The elementary process of the decay is 
\begin{align}
 b \to s + \phi^\ast \to s + A' + A', \label{eq:b-decay}
\end{align}
where $b$ and $s$ denote bottom and strange quark, respectively, and $\phi^\ast$ off-shell dark Higgs boson. In the loop, all up-type quarks 
propagate and contribute to the decay. However, the decay process is dominated by the top quark loop due to the large Yukawa coupling. 
Since we consider the situation where the dark Higgs boson is lighter than $W$ boson and top quark $t$, the off-shell decay of 
$b \to s + \phi^\ast$ can be described by an effective Lagrangian \cite{Grinstein:1988yu, Chivukula:1988gp} as 
\begin{align}
\mathcal{L}_{\Delta B=1}^{\mathrm{eff}} = &\frac{3 \alpha_{\mathrm{em}}}{32 \pi s_W^2} V_{tb} 
\frac{m_t^2}{m_W^2} V_{ts}^\ast \frac{c_\alpha h + s_\alpha \phi}{v} 
 m_b \overline{s} \left( 1 + \gamma^5 \right) b + \mathrm{H.c.} , \label{eq:effective-lag}
\end{align}
where $\alpha_{\mathrm{em}}$ is the fine structure constant and $V_{tb},~V_{ts}$ are the Cabibbo-Kobayashi-Maskawa matrix element 
\cite{Cabibbo:1963yz, Kobayashi:1973fv, Workman:2022ynf}.
The masses of top, bottom quark and $W$ boson are denoted as $m_t$, $m_b$ and $m_W$, respectively.

Using the effective Lagrangian eq.~\eqref{eq:effective-lag} and the gauge-scalar interaction, the decay amplitude of 
the process \eqref{eq:b-decay} can be given by 
\begin{align}
\label{amp_3-body_decay}
M \left( b \to s A' A' \right) &= \left( \frac{3 \alpha_{\mathrm{em}}}{32 \pi s_W^2} V_{tb} \frac{m_t^2}{m_W^2} V_{ts}^\ast \frac{s_\alpha}{v} m_b \right) \left( i \frac{m^2_{A'}}{v_\Phi} c_\alpha \right) \nn \\
&\times \overline{u} \left( p_1 \right) \left( 1 + \gamma_5 \right) u \left( k \right) \frac{i}{s_{23} - m_\phi^2 + i \mathrm{Im}( \Pi(s_{23}) ) } \epsilon_1^{* \mu} \left( p_2 \right) \epsilon_{2 \mu}^* \left( p_3 \right),
\end{align}
where four momentum assignment to each particle is shown in figure~\ref{fig:B-meson-decay} and  $s_{23} \equiv (p_2+p_3)^2$. 
The Dirac spinor with four momentum $p$ is denoted by $u(p)$, and similarly the polarization vector of dark photon is 
$\epsilon_i^\mu(p),(i=1,2)$, respectively. In the propagator of the dark Higgs boson, $\Pi(s_{23})$ represents the sum of 
one-particle irreducible amplitudes for two-point function of $\phi$. 
When the propagator has narrow peak around $m_\phi$, the imaginary part can be given by $m_\phi \Gamma(m_\phi^2)$. However, in our case, the imaginary part can be comparable to $m_\phi$ due to the decay of $\phi \to A'A'$, in which 
the longitudinal polarization enhances the amplitude. Thus, the $s_{23}$ dependence of $\Pi$ should be kept. 
In order to evaluate the imaginary part numerically,  we replace it with $\sqrt{s_{23}}~ \Gamma_\phi(s_{23})$.

The differential decay width with respect to $s_{23}$ is given by
\begin{align}
\frac{d}{d s_{23}} \Gamma \left( b \to s A'A' \right) &= \frac{9(g'\alpha_{\mathrm{em}} s_\alpha c_\alpha )^2 m_{A'}^2 m_b^3}{2^{16} \pi^5 s_W^4 v^2} \left| V_{tb} \right|^2 \left| V_{ts} \right|^2 \frac{m_t^4}{m_W^4}  \left( 1 - \frac{s_{23}}{m_b^2} \right)^2 \nn \\
&\times \sqrt{1 - \frac{4 m_A'^2}{s_{23}}} 
\frac{1}{\left( s_{23} - m_\phi^2 \right)^2 + s_{23} (\Gamma_\phi(s_{23}))^2 } \left[ 2 + \left( \frac{s_{23}}{2 m_{A'}^2} - 1 \right)^2 \right], \label{eq:diff-width}
\end{align}
where the second term in the last parenthesis represents the longitudinal polarization.
The partial decay width can be obtained by integrating over $s_{23}$ from $4 m_{A'}^2$ to $m_b^2$. 
It is noticed that the contribution from $K$ meson decay is negligible due to the small CKM angle or Yukawa coupling suppression of charm quark.

%%%%%%%%%%%%%%%%%%%%%%%%%%%%%%%%%%%%%%%%%%%%%%%%%%%%%%%%
\subsection{Expected number of signal events at FASER}
%%%%%%%%%%%%%%%%%%%%%%%%%%%%%%%%%%%%%%%%%%%%%%%%%%%%%%%%
\begin{table}[t]
\begin{tabular}{|c|c|c|c|c|} \hline
 \hspace{2cm} & ~~~$L_{\rm min}$~(m)~~~ & ~~~$L_{\rm max}$~(m)~~~ & ~~~$R$~(m)~~~ & ~~~$\mathcal{L}$~(ab$^{-1}$)~~~ \\ \hline \hline 
 FASER  & 478.5 & 480 & 0.1 & 0.15 \\ \hline
 FASER~2 & 475 & 480 & 1.0 & 3.0 \\ \hline
\end{tabular}
\caption{
Dimension of the FASER and FASER2 detector and integrated luminosity used for this study. 
$L_{\rm min}$ and $L_{\rm max}$ are the distance to the front and rear end of the detector from the IP, respectively. 
$R$ is detector radius, respectively. 
$\mathcal{L}$ is the integrated luminosity.
}
\label{tab:faser-dimension}
\end{table}
%%%%%%%%%%%%%%%%%%%%
Next, we show the number of events at the FASER and future upgrade FASER2 experiment. The FASER detector is located along the beam axis of LHC and approximately 480 m away from the proton interaction point of ATLAS. 
It is cylindrical shape of $10$ cm radius and $1.5$ m long.
Three tracking spectrometers with $0.57$ T dipole magnet are installed so that the charged particles can be identified. 
The dark photon can decay into a pair of oppositely charged particles, such as $e^\pm, \mu^\pm, \pi^\pm$, and $K^\pm$, when the final states are kinematically allowed.
These charged particles leave tracks in the FASER detector while the dark photon itself does not. Such an event with only two charged tracks is regarded as the signal of the dark photon decay.
Details of the detector can be found in ref.~\cite{FASER:2022hcn}. The dimension of FASER2 detector is now under discussion \cite{Feng:2022inv,Anchordoqui:2021ghd,Salin:2024jjm}. We employ the same setup studied in ref.~\cite{Araki:2020wkq}, i.e. cylindrical shape detector with 1 m radius and 5 m long. 
Dimension of the detector and integrated luminosity for our study are summarized in table~\ref{tab:faser-dimension}.

The dark photon produced from $B$ meson decay propagates to the FASER/FASER2 detector, and then decays into the pair of the SM charged particles such as $e^\pm, \mu^\pm, \pi^\pm$, and $K^\pm$. In the mass region of our interest, the branching ratios into neutral particles are negligible\footnote{
Dark photon can decay into a pair of $K^0$ mesons as well, and the decay is enhanced due to the narrow resonance with a $\phi$ meson.
The inclusion of the decay, however, only affects the sharp dip at $m_{A'}=1$ GeV in figures \ref{fig:sensitivity1} and \ref{fig:sensitivity2}.
Moreover, its decay branching ratio is smaller than that of $K^\pm$ mesons.
Thus, we ignore the decay into $K^0$ mesons in this work.
}. 
We treat the decay branching ratio as unity. 
The expected number of dark photon decay is calculated by 
\begin{align}
\label{eq:num-of-event}
   N
   &= \mathcal{L} \int dp_B d\theta_B 
   \frac{d\sigma_{pp \to B}}{dp_B d\theta_B }~
   \int_{4m_{A'}^2}^{m_b^2} d s_{23}
   \left[ \frac{d}{d s_{23}} {\rm Br}(B \to X_s A'A') \right] 
   \left( \mathcal{P}_{A'}^{\rm det}(\bm{p}_2) + \mathcal{P}_{A'}^{\rm det}(\bm{p}_3) \right)~,
\end{align}
where the integrated luminosity is written as $\mathcal{L}$, the momentum and angle of each $B$ meson are denoted as $p_B$ 
and $\theta_B$, respectively. The differential cross section of $B$ meson is taken from the data file in FORESEE package \cite{Kling:2021fwx}.
The differential decay branching ratio is given by
\begin{align}
\frac{d}{d s_{23}} {\rm Br}(B \to X_s A'A') \equiv \frac{1}{\Gamma_b} \frac{d}{d s_{23}} \Gamma(b \to s A'A'),
\end{align}
where $\Gamma_b$ is the full width of $b$ quark.
The probability that $A'$ decays inside the detector with momentum $\bm{p}$ is written 
by $\mathcal{P}_{A'}^{\rm det}(\bm{p})$ which is given by
\begin{align}
 \mathcal{P}_{A'}^{\rm det}(\bm{p})=\left( 
 e^{-L_{\mathrm{min}}/\beta\gamma\tau_{A'}} - e^{-L_{\mathrm{max}}/\beta\gamma\tau_{A'}}
 \right) \Theta\left(R - L_\mathrm{min} \tan\theta \right),
\end{align}
where $\beta\gamma$ is the Lorentz factor corresponding to the momentum $|\bm{p}|$, and $\theta$ is the angle of the momentum to the 
beam direction. The last term $\Theta(x)$ is the step function. The distance to the front (rear) end of the detector from the IP is denoted as $L_{\mathrm{min(max)}}$, and the radius of the 
detector is $R$. These parameters are given in table~\ref{tab:faser-dimension}.
%

%%%%%%%%%%%%%%%%%%%%%%%%%%%%%%%%
\section{Constraints} \label{sec:constraints}
%%%%%%%%%%%%%%%%%%%%%%%%%%%%%%%%
The dark photon and dark Higgs boson have been searched by beam dump, collider and neutrino experiments so far,  and  
tight constraints are placed separately on the parameter space of those particles \cite{Feng:2022inv}. 
These constraints should be applied to the dark photon and dark Higgs boson in our model. 
In addition, there are other constraints which arise from the interactions of the dark photon with the dark Higgs boson. 
One of such constraints is the decay of the SM Higgs boson into the dark Higgs bosons and dark photons through the scalar mixing. 
The other one is the perturbative unitarity constraint on the dark photon scattering, when we assume that the perturbative 
expansion is valid at least at the scale of experiments. In this section, we derive these constraints in our dark photon model.

%%%%%%%%%%%%%%%%%%%%%%%%%%%%%%%%
\subsection{Invisible decay of SM Higgs boson}
%%%%%%%%%%%%%%%%%%%%%%%%%%%%%%%%
The SM Higgs boson can decay into a dark photon pair through the scalar mixing when the dark photon is light enough. 
Since the dark photon is long-lived in the parameter space of our interest, the dark photons will escape detectors. 
Then, such a decay will be regarded as invisible decay of the SM Higgs boson in experiments. 
The partial width of this decay can be significantly enhanced due to the longitudinal mode of the dark photon, resulting in a large branching 
ratio of the invisible decay of the SM Higgs boson. 

The partial decay widths of $h \to A'A'$ and $h \to \phi\phi$ are given by
\begin{subequations}
    \begin{align}
\Gamma(h \to A'A') &= \frac{g'^2 \sin^2\alpha}{8 \pi}\frac{m_{A'}^2}{m_h} \sqrt{1-\frac{4m_{A'}^2}{m_h^2}} 
\left( 2 + \frac{m_h^4}{4 m_{A'}^4} \left(1 - \frac{2 m_{A'}^2}{m_h^2} \right)^2 \right), \\
\Gamma(h \to \phi \phi) &= \frac{{g'}^2 c_\alpha^6 \tan^2 2\alpha}{128 \pi} \frac{m_h^3}{m_{A'}^2} \sqrt{1-\frac{4 m_{A'}^2}{m_h^2}},
\end{align}
\end{subequations}
where we used the term proportional to $\lambda_{H\Phi} v c_\alpha^3$ and neglected other terms
due to the suppression by $m_\Phi/m_{A'}$ and $\alpha^4$. We neglected the partial decay width of $h \to \phi\phi\phi$ because of the smaller coupling constants and phase space suppression.
The branching ratio of the invisible decay of the SM Higgs boson is given by
\begin{align}
\mathrm{Br}(h \to \mathrm{invisible}) = \frac{\Gamma(h \to A'A') + \Gamma(h \to \phi \phi) + \Gamma(h \to 4\nu)}{\Gamma_h + \Gamma(h \to A'A') + \Gamma(h \to \phi \phi)}, \label{eq:invisible-br}
\end{align}
where $\Gamma_h = 4.07$ MeV is the total decay width predicted by the SM \cite{Workman:2022ynf}, and 
$\Gamma(h \to 4\nu) = 1.2 \times 10^{-3} \Gamma_h $ is the partial decay width into $\nu \bar{\nu}\nu \bar{\nu}$ \cite{LHCHiggsCrossSectionWorkingGroup:2016ypw,Workman:2022ynf}.
The current experimental upper bounds on the invisible decay branching ratio of the SM Higgs boson are placed by ATLAS \cite{ATLAS:2023tkt} and 
CMS \cite{CMS:2022qva, CMS:2023sdw}. 
ATLAS sets more stringent bound at $95$\% C.L. combined Run 1 and 2 data as  
\begin{align}
\mathrm{Br}^{\mathrm{comb}}_{\mathrm{inv}} < 0.107. \label{eq:invisible-exp}
\end{align}
This bound will be improved at the High-Luminosity LHC (HL-LHC) 
as
\begin{align}
    \mathrm{BR}_{\rm inv}^{\rm HL-LHC} < 0.025.~~~(95\% \mathrm{C.L.})
\end{align}
We use these current and expected upper limits in our analysis.

%%%%%%%%%%%%%%%%%%%%%%%%%%%%%%%%
\subsection{Perturbative unitarity}
%%%%%%%%%%%%%%%%%%%%%%%%%%%%%%%%
Requiring the perturbative expansion of amplitudes to be valid up to the LHC beam energy scale, the dark Higgs boson mass can be bound by 
the dark photon mass and gauge coupling. This is because the dark Higgs boson is the origin of the dark photon mass, and known 
as perturbative unitarity bound \cite{Lee:1977yc, Lee:1977eg}. 

The constraint can be obtained from the amplitudes of $A' A' \to A'A',~A'A' \leftrightarrow \phi \phi,~\phi \phi \to \phi \phi$ and 
$A' \phi \to A' \phi$ scatterings. 
In the center of mass frame of the initial particles, the scattering amplitude $M$ can be written by using partial wave expansion as
\begin{align}
M = \sum_{l=0}^\infty 8 \pi \frac{E_{\rm CM}}{|\bm{p}|} (2l + 1) a_l(|\bm{p}|) P_l(\cos\theta),
\end{align}
where $E_{\rm CM}$ and $\bm{p}$ denote the energy and momentum of the initial particle, respectively, and $\theta$ is the scattering angle of a 
final particle with respect to $\bm{p}$.  The Legendre polynomial is denoted as $P_l(\cos\theta)$. The expansion coefficient $a_l(|\bm{p}|)$ is 
restricted so that the perturbative unitarity is maintained. The constraint for the S-wave coefficient is given by
\begin{align}
\left| \mathrm{Re}(a_0(\bm{p})) \right| < \frac{1}{2}.
\end{align}
The coefficients for the scatterings of $A'$ and $\phi$ are read from the scattering amplitudes using eqs.~\eqref{eq:lag-gauge-scalar} 
and \eqref{eq:lag-scalar},
\begin{subequations}
\begin{align}
&a_0(A' A' \to A'A') = -\frac{3}{16\pi}\frac{g'^2 m_\phi^2}{m_{A'}^2}, \\
&a_0(A'A' \leftrightarrow \phi \phi) = \frac{1}{16\pi} \left( 2g'^2 + g'^2 \frac{m_\phi^2}{m_{A'}^2}  - 4 g'^2 \log\left( \frac{s}{m_{A'}^2} \right)\right),\\
&a_0(\phi \phi \to \phi \phi) = -\frac{3}{16\pi}\frac{g'^2 m_\phi^2}{m_{A'}^2},\\
&a_0(A' \phi \to A' \phi) = \frac{1}{16\pi} \left( g'^2 - g'^2 \frac{m_\phi^2}{m_{A'}^2}  -2 g'^2 \log\left( \frac{s}{m_{A'}^2} \right)\right),
\end{align}
\end{subequations}
where $s$ is the squared center-of-mass energy.
It should be noticed that the $\log(s)$ term cannot be neglected when $m_{A'}$ is much smaller compared with $\sqrt{s}$, in particular for large $g'$.
Then, diagonalizing the matrix of the coefficients in basis of initial states $A'A'$, $\phi\phi$ and $A'\phi$, the constraint is given by the largest eigenvalue, 
$|a_{0,\mathrm{max}}|$ 
as
\begin{align}
|a_{0,\mathrm{max}}| < \frac{1}{2}. \label{eq:unitarity-bound}
\end{align}
As we will see in the next section, the perturbative unitarity bound can be tighter in heavy $\phi$ spectrum than the invisible decay of the SM Higgs boson.

%%%%%%%%%%%%%%%%%%%%%%%%%%%%%%%%
\section{Sensitivity region} \label{sec:sensitivity-region}
%%%%%%%%%%%%%%%%%%%%%%%%%%%%%%%%
We now show the expected sensitivity region of dark photon decays in $m_{A'}\,\mathchar`-\,\varepsilon$ plane, where $\varepsilon$ is defined just below eq.~\eqref{eq:dp-Jem}.
%\TS{erased: \sout{, for the FASER2 setup}}. 
The expected number of events is calculated by using eqs.~\eqref{eq:diff-width} and \eqref{eq:num-of-event}.
As was concluded in ref.~\cite{FASER:2023tle}, main sources of background at the FASER experiment are neutrinos and neutral hadrons, and a total background is estimated as $(2.3\pm2.3)\times 10^{-3}$ events for 27 fb$^{-1}$.
Although the background is suppressed enough to assume background free for FASER, it might become non-negligible level for the FASER2 setup if the background increases at this rate as the detector volume and the luminosity increase.
Nevertheless, the background could be reduced with more complicated trigger techniques and an increase of neutrino interaction data; At this moment, it is difficult to precisely anticipate the background.
Thus, we also assume background free for the FASER2 setup as with the previous studies \cite{Feng:2017uoz,FASER:2018eoc,Araki:2020wkq} and compare our new results with the previous ones.
\begin{figure}[t]
\begin{tabular}{cc}
\hspace{-10mm}
    \includegraphics[scale=0.35]{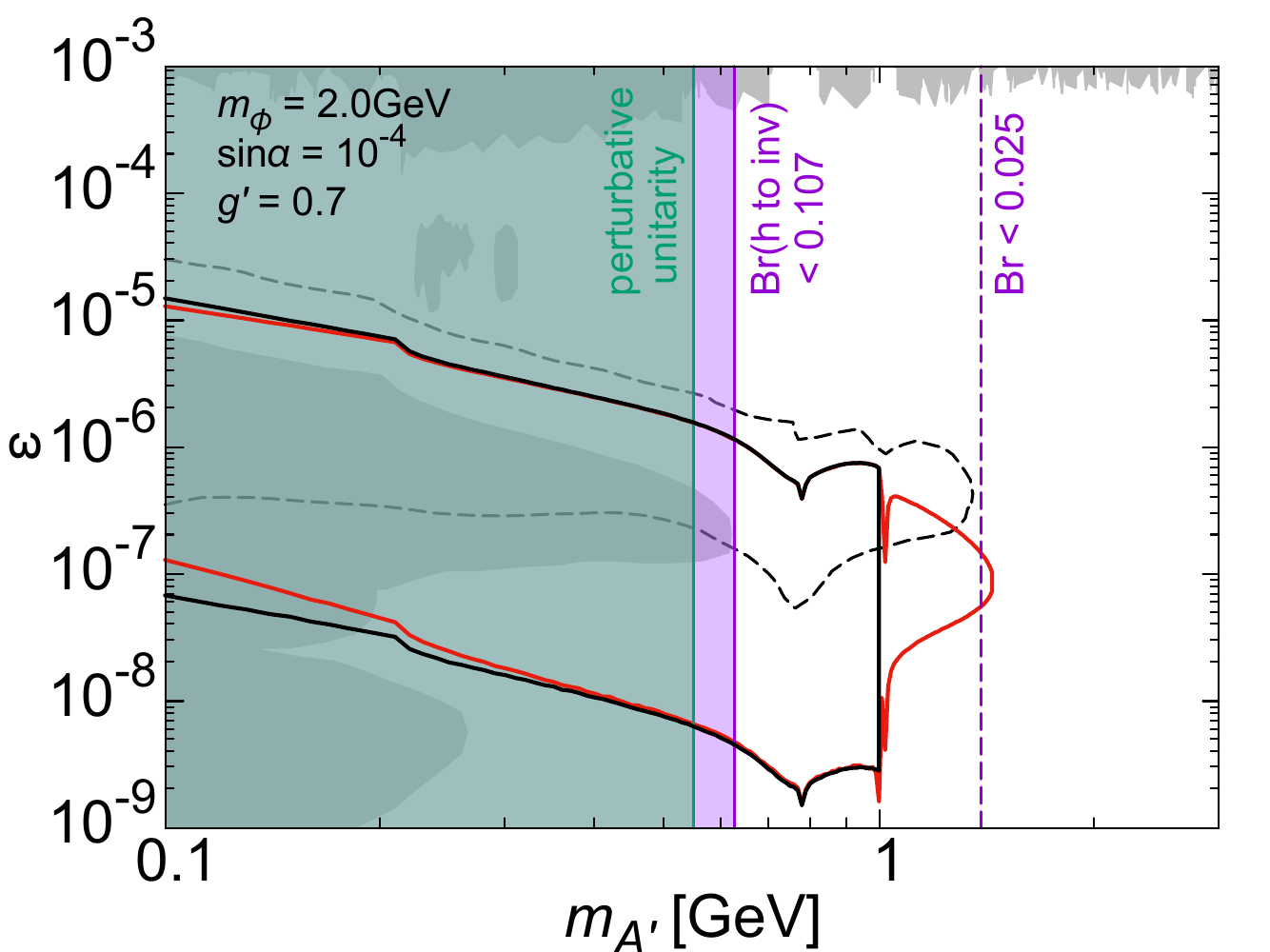} &
    \includegraphics[scale=0.35]{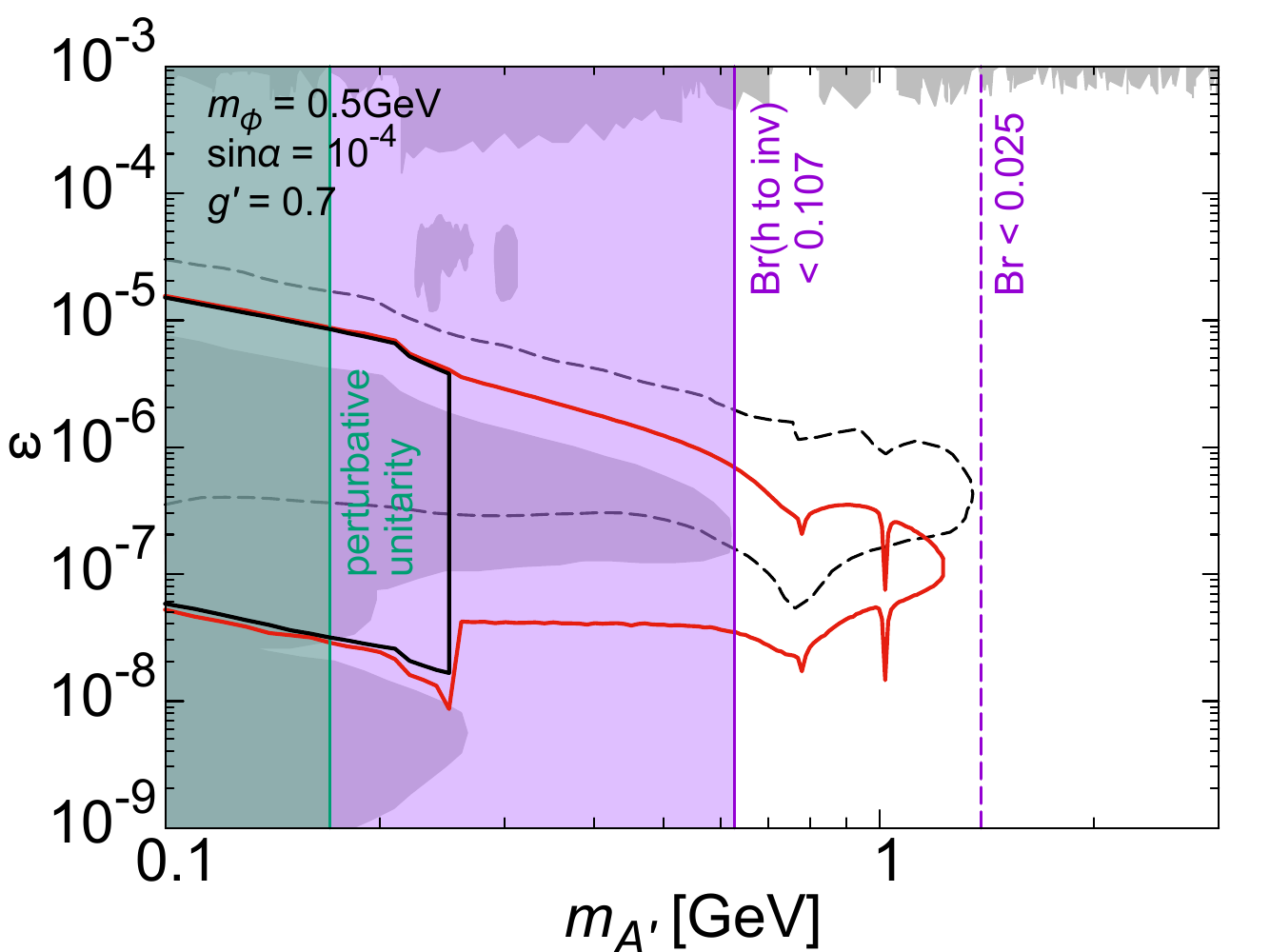}
\end{tabular}
    \includegraphics[scale=0.35]{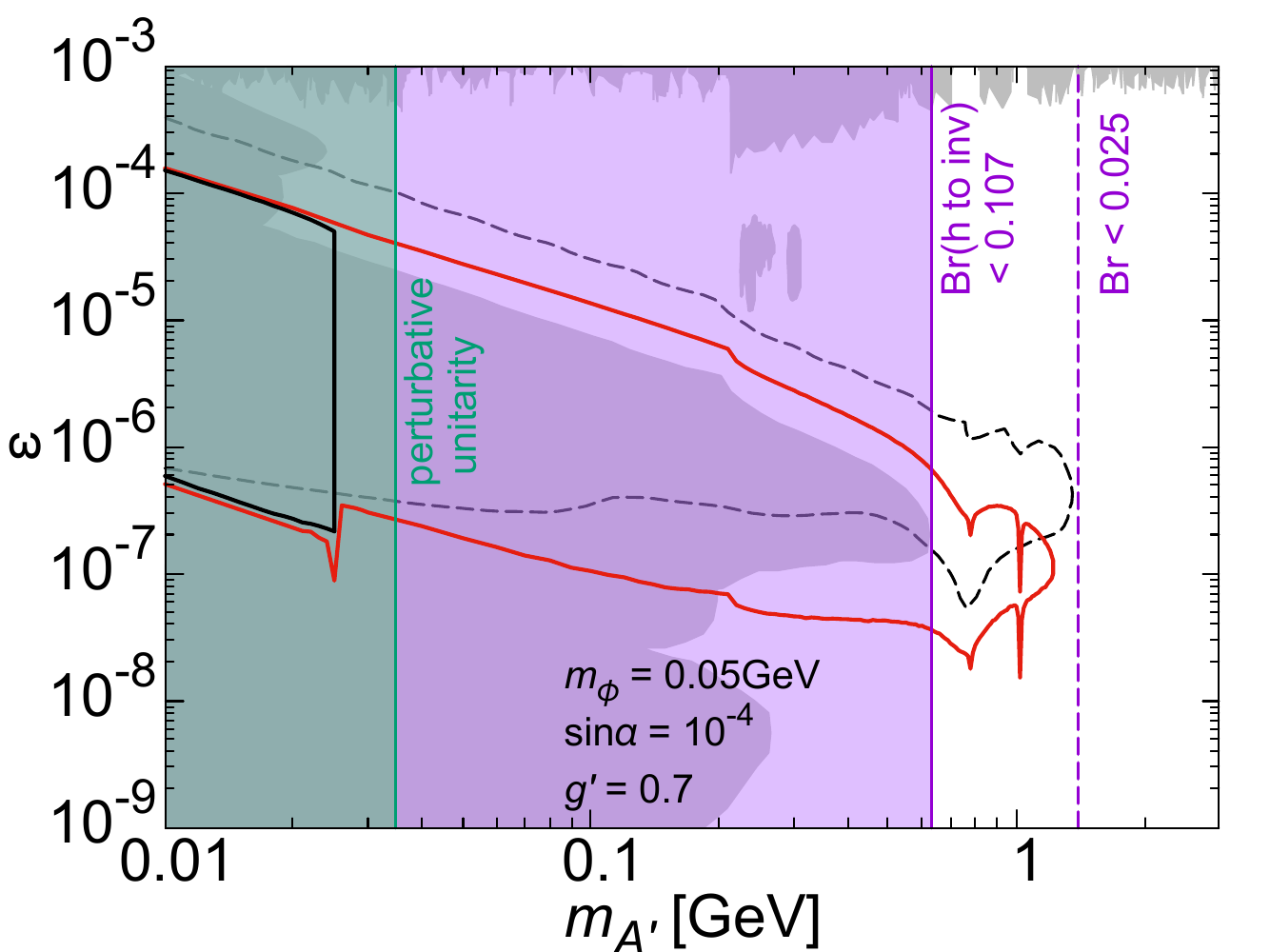}
    \caption{The $95$\% C.L. sensitivity contours for the dark photon decays for $m_\phi=2$ (top left), $0.5$ (top right) and $0.05$ (bottom) GeV. 
    The gauge coupling is taken to be $g'=0.7$. 
    The red curves represent our results which include both the on-shell and off-shell dark Higgs state, while the black solid curves are only for the on-shell dark Higgs state. 
    Green and purple colored regions are excluded by the perturbative unitarity and invisible decay of the SM Higgs boson constrains, respectively.  Details can be found in the text.}
\label{fig:sensitivity1}
\end{figure}
In eq.~\eqref{eq:diff-width}, the on-shell state as well as the off-shell state of the dark Higgs boson is  
automatically included. As representative examples, the parameters are chosen as $g'=0.7$ and $0.5$ for $m_\phi = 2,~0.5$ and $0.05$ GeV,\footnote{For $g'<0.7$, the log(s) term 
as well as the constant $g'^2$ term can be safely neglected for our choice of parameters. However, for $g'>0.7$, these terms becomes 
comparable or larger than $g'^2 m_\phi^2/m_{A'}^2$.} while the scalar mixing is fixed to be $\alpha = 10^{-4}$. We found that, for the 
FASER setup, the expected number of events at 95\% C.L was not found due to the smaller detector size and luminosity. Thus, in the following, we only show the results for the FASER2 setup.

Figure~\ref{fig:sensitivity1} shows the sensitivity region for $g'=0.7$ at $95$\% C.L. which corresponds to $3$ signal events with null background. 
The mass of the dark Higgs boson is taken to be $2$ (top left), $0.5$ (top right) and $0.05$ (bottom) GeV, respectively. 
The red solid curve in each panel represents our result which includes the on-shell and off-shell dark Higgs state, while the black solid one represents 
the sensitivity region only for the on-shell decay of the dark Higgs boson calculated by following ref.~\cite{Araki:2020wkq}. 
Below the kinematical threshold $m_{A'} < m_\phi/2$, the dark Higgs boson is short-lived as was shown in figure~\ref{fig:DecayLengthOfDarkHiggs}, and promptly decays into the two dark photons. For this spectrum,   
the on-shell decay of $B \to X_s + \phi$ followed by $\phi \to A'A'$ dominates the branching ratio of $B \to X_s A'A'$ 
due to the narrow resonance at $s_{23}=m_\phi^2$ in eq.~\eqref{eq:diff-width}. 
Thus, the black and red curves are almost the same in this mass region.
The slight difference can be seen between the black and red curves below the threshold especially for light $m_{A'}$. 
This is because the narrow width approximation is not valid for such spectrum. Due to the longitudinal enhancement of 
$\phi \to A'A'$ decay, the $\phi$ decay width is comparable to its mass. 
The black dashed curve shows $95$\% C.L. sensitivity region of the so-called vanilla dark photon model without dark Higgs boson for comparison \cite{Feng:2022inv}. The grey regions are excluded by past experiments and observation \cite{Kling:2021fwx} (references therein). The purple vertical lines indicate the lower limit on 
the dark photon mass obtained from the invisible decay of the SM Higgs boson, eq.~\eqref{eq:invisible-br},  
corresponding to $\mathrm{Br}(h \to \mathrm{inv.}) = 0.107$ (solid) and $0.025$ (dashed), respectively. 
The purple colored region is excluded by the present experimental result \cite{ATLAS:2023tkt}.
The green colored region is also excluded by the perturbative unitarity constraint in eq.~\eqref{eq:unitarity-bound}. For the unitarity constraint, we set $\sqrt{s}=10~\mathrm{TeV}$, which is about the center of mass energy in ATLAS.

\begin{figure}[t]
\begin{tabular}{cc}
\hspace{-10mm}
    \includegraphics[scale=0.35]{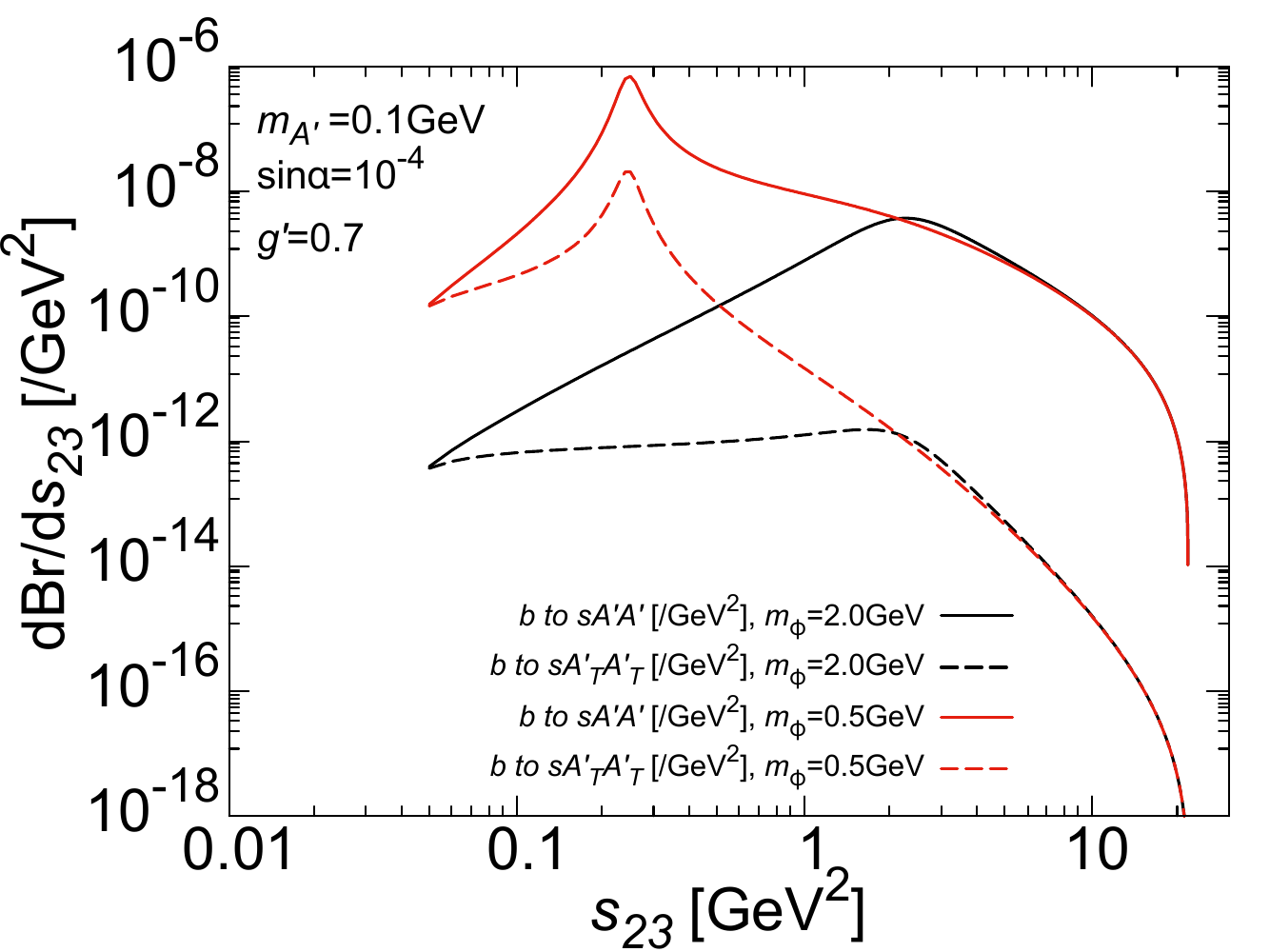} &
    \includegraphics[scale=0.35]{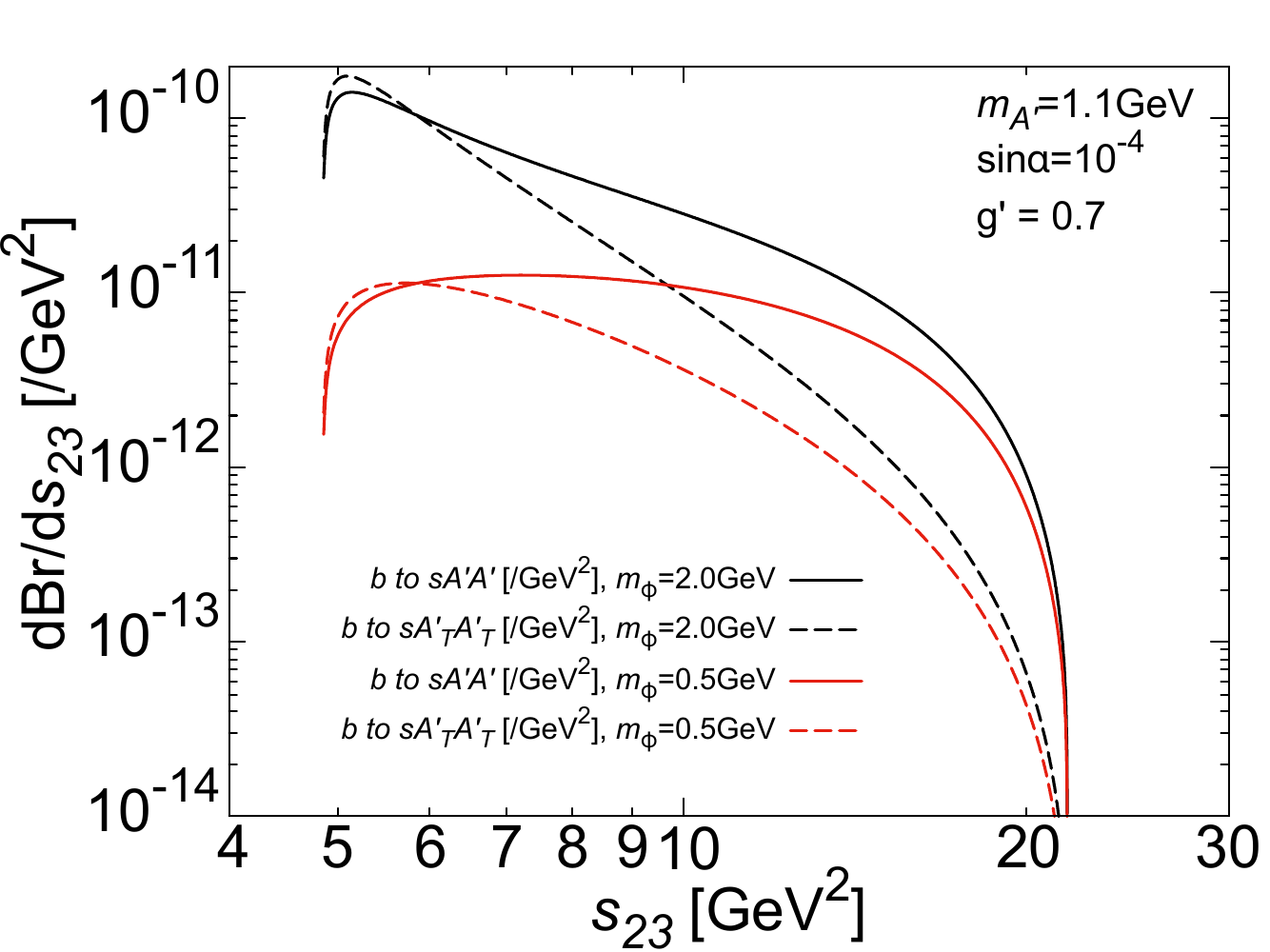}
\end{tabular}
    \caption{Differential branching ratio with respect to $s_{23}$. The dark photon mass is taken to $0.1$ (left) and $1.1$ GeV. 
    Black and red curves correspond to the $m_\phi=2$ and $0.5$ GeV in both panel. The solid curve represents the decay into both 
    the longitudinal and transverse contribution of the dark photon while the dashed one represents only transverse component.}
\label{fig:diff-br}
\end{figure}
From the figure, one can see that the sensitivity regions cover the wide area of the parameter space which cannot be explored 
by the on-shell decay of the dark Higgs boson.
In each panel, the sensitivity regions reach to $m_{A'} \simeq 1.2$ GeV which is almost independent of $m_\phi$.  
This behavior is attributed to the follows facts. In the region where $2m_{A'} > m_\phi$, 
$s_{23}$ is always larger than $m_\phi^2$. The propagator in eq.~\eqref{eq:diff-width} loses $m_\phi$ 
dependence and behaves as $1/s_{23}^2$ for large $s_{23}$. Such a  propagator is canceled by the longitudinal contributions of the dark photon 
in the $b$ decay. This feature is shown in figure~\ref{fig:diff-br} and explained later. 
For heavy $m_\phi$, the perturbative unitarity constraint is more stringent than the invisible decay of the SM Higgs boson. It has already excluded the dark photon mass 
$m_{A'} < 0.8$ GeV.\footnote{For $g'=1$, the lower limit on $m_{A'}$ from the perturbative unitarity constraint is $\mathcal{O}(10)$ GeV. This bound 
is mainly determined by the $\log(s)$ term. In such a case, higher order processes should be taken into account.} 
On the other hand, for light $m_\phi$, the constraint on the invisible decay of the SM Higgs boson has already excluded the light dark photon from 
the on-shell decay of the dark Higgs boson. Therefore, the off-shell decays of the dark Higgs bosons should be taken into account for dark photon search. 
Figure~\ref{fig:diff-br} is the comparison of the differential branching ratios of $b \to s A'A'$ with that for the transverse components $b \to s A'_T A'_T$ 
with respect to $s_{23}$.
The mass of the dark Higgs boson is chosen as $m_\phi = 2$ (black) and $0.5$ (red) GeV, and the dark photon mass is 
taken to $0.1$ (left) and $1.1$ GeV (right), respectively. The solid and dashed curves represent the total and transverse contribution, respectively. Thus, the difference between both curves is the longitudinal contribution. The panels show that the longitudinal contribution dominates 
the differential branching ratio except for the lower and upper kinematical thresholds. 
\begin{figure}[t]
\begin{tabular}{cc}
\hspace{-10mm}
    \includegraphics[scale=0.35]{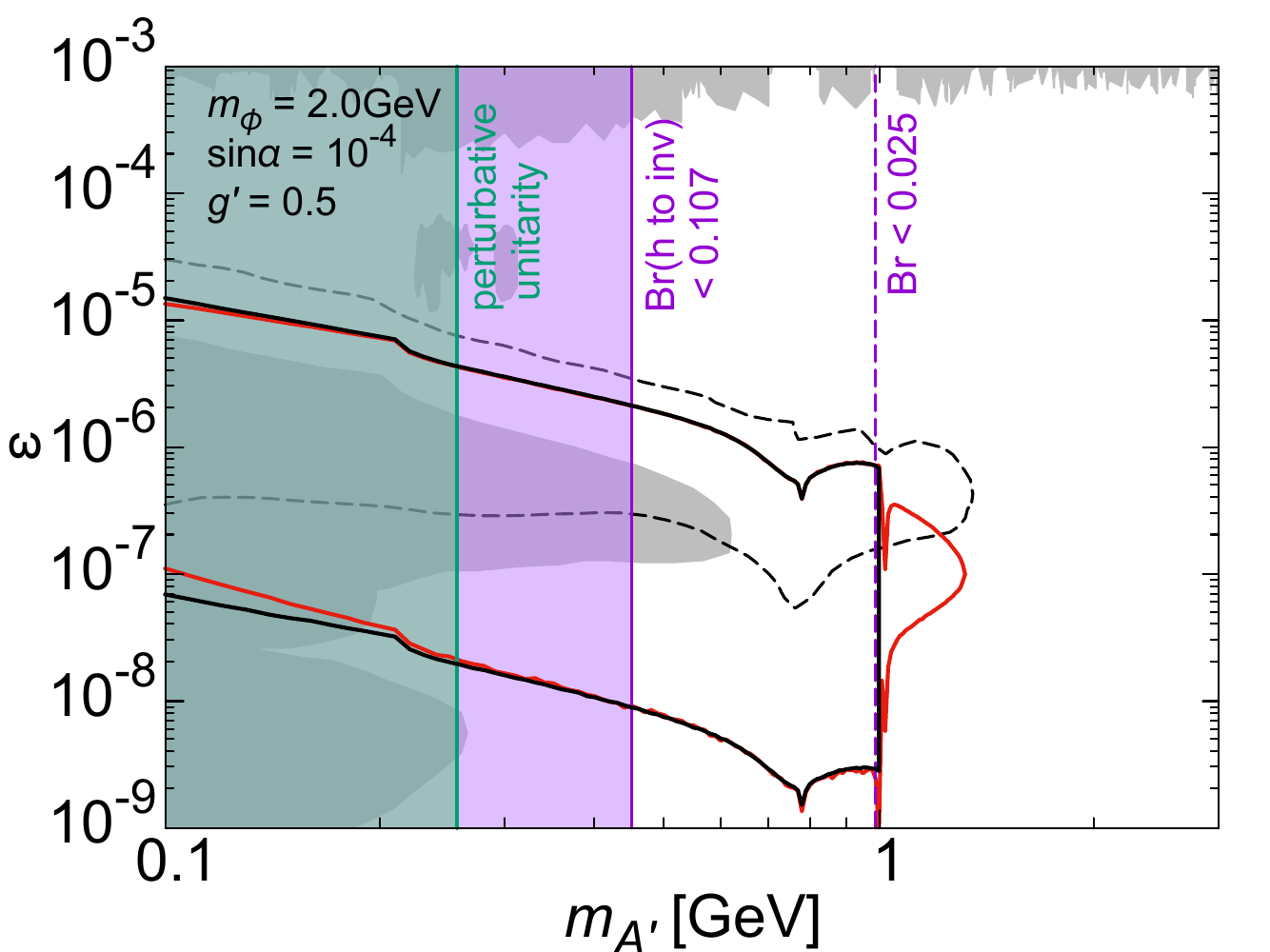} &
    \includegraphics[scale=0.35]{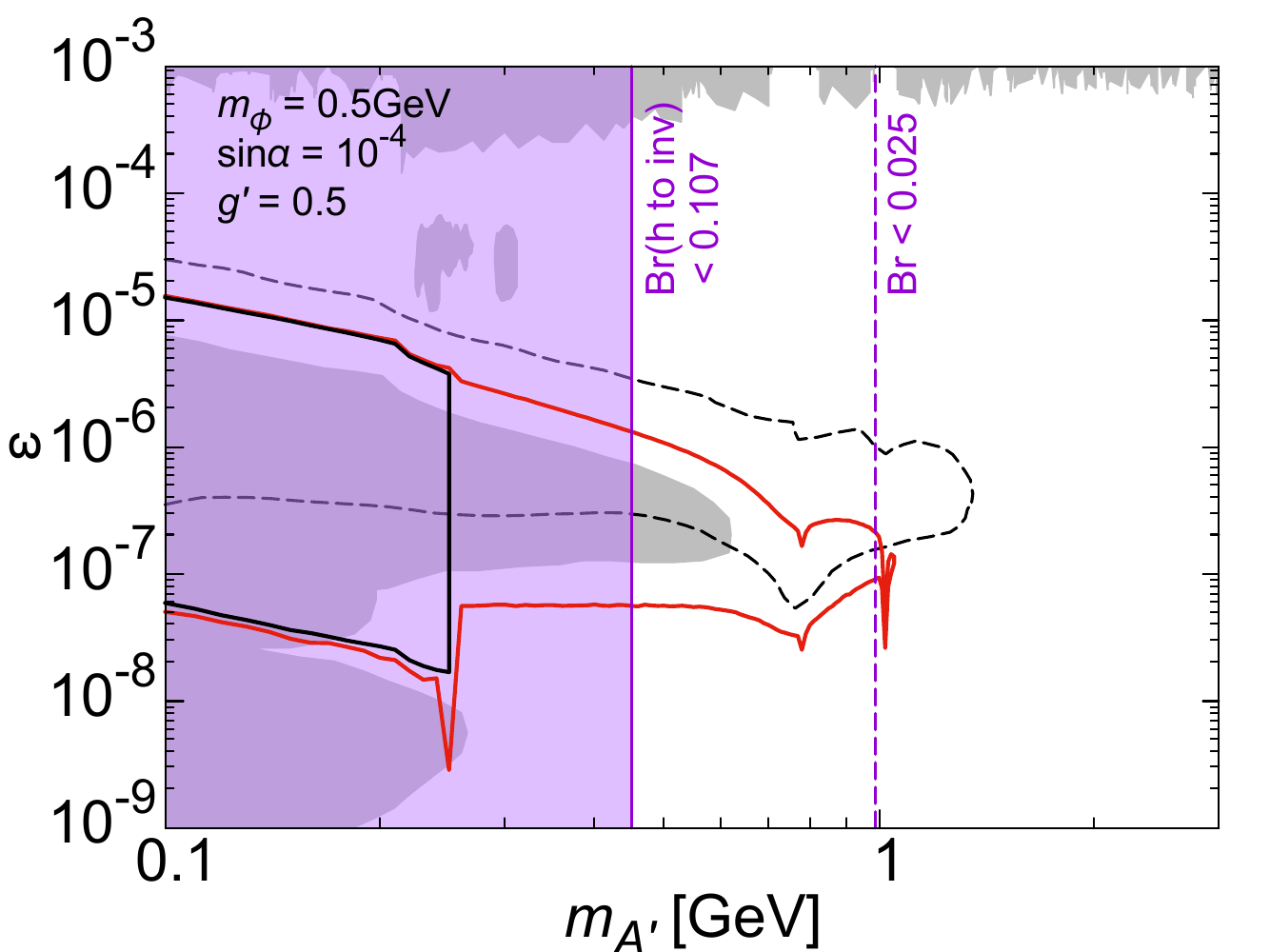}
\end{tabular}
    \includegraphics[scale=0.35]{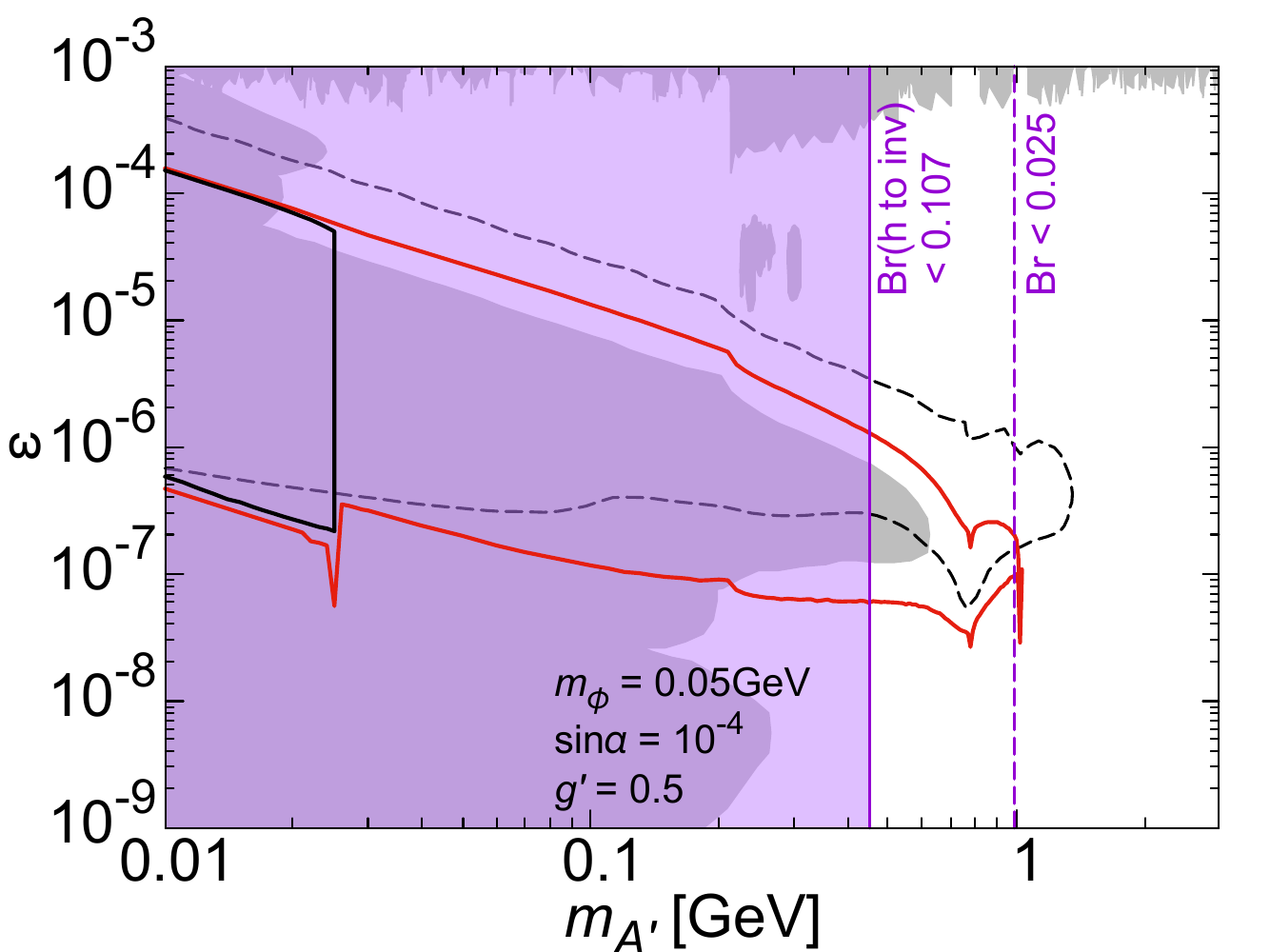}
    \caption{The same plot as Figure \ref{fig:sensitivity1} for $g'=0.5$.}
\label{fig:sensitivity2}
\end{figure}
In the left panel, the dark photon is light so that it can be produced by the on-shell decay of the dark Higgs boson. Such a decay can be seen 
as a resonance peak in the differential branching ratio. For $m_\phi = 2$ GeV, the resonance peak is broad due to the large $\Gamma_\phi(s_{23})$. 
On the other hand, in the right panel, the dark photon is only produced via the off-shell decay of the dark Higgs boson. Right panel shows that 
the total differential branching ratios slowly decrease than those only with the transverse contribution do. This is because the $s_{23}$ dependence compensates between the propagator and the longitudinal contribution. 

Figure~\ref{fig:sensitivity2} is the same plots as figure~\ref{fig:sensitivity1} for $g'=0.5$. 
The sensitivity regions are narrower since the expected number of events is proportional to $g'^2$. Nonetheless, the sensitivity region from the off-shell decay of the dark Higgs boson appears in smaller $\epsilon$ region than 
that from proton bremsstrahlung and meson decay production. 
For $g' < 0.5$, the sensitivity region from the off-shell decay of the dark Higgs boson coincides with the on-shell one.

%%%%%%%%%%%%%%%%%%%%%%%%%%%%%%%%
\section{Conclusion} \label{sec:conclusion}
%%%%%%%%%%%%%%%%%%%%%%%%%%%%%%%%
We have considered a dark photon model with the dark Higgs boson which gives a mass to the dark photon through spontaneous 
symmetry breaking of the dark symmetry. We studied the off-shell contribution of the dark Higgs boson to dark photon production, and 
derived the expected sensitivity reach of dark photon signal at the FASER2 experiment. 

We have shown that the off-shell dark Higgs boson contribution for the dark photon production is important at the FASER experiment. 
Even for the light dark Higgs boson, the off-shell process is unsuppressed due to the longitudinal component of the dark photon, and the sensitivity region  
extends to heavier dark photon mass. We also derived the constraint from the invisible decay of the SM Higgs boson and perturbative unitarity. The perturbative 
unitarity constraint is stringent in heavy mass region of the dark Higgs boson due to the longitudinal enhancement. For lighter dark Higgs boson, the constraint on the invisible decay of the SM Higgs boson excluded the on-shell production of the dark photon for $g'>0.5$ as shown in figures~\ref{fig:sensitivity1} and \ref{fig:sensitivity2}. Thus, the off-shell decay of the dark Higgs boson should be taken into account for the dark photon search.

\section*{Note added}
\begin{figure}[t]
\begin{tabular}{cc}
\hspace{-10mm}
    \includegraphics[scale=0.35]{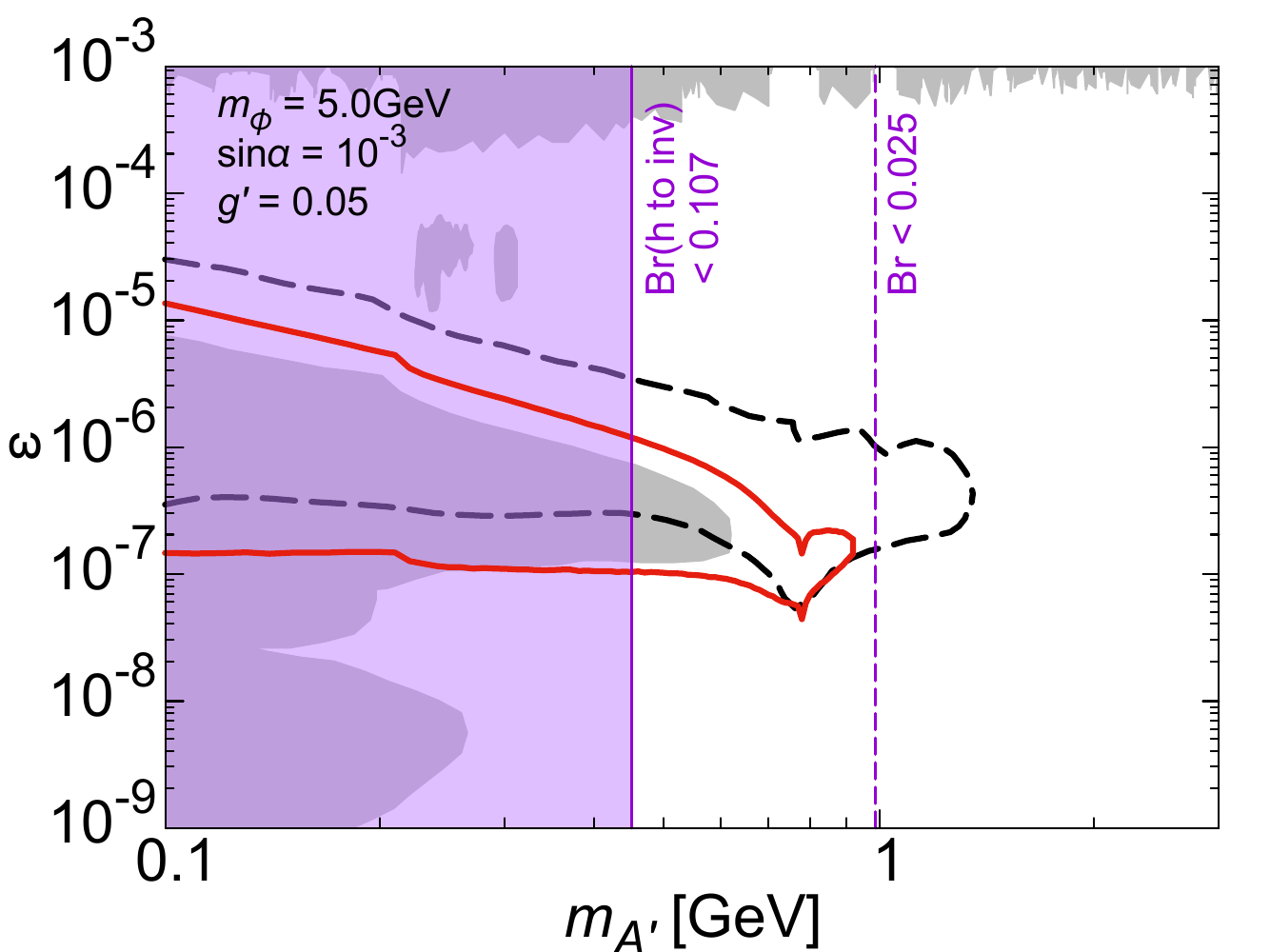} &
    \includegraphics[scale=0.35]{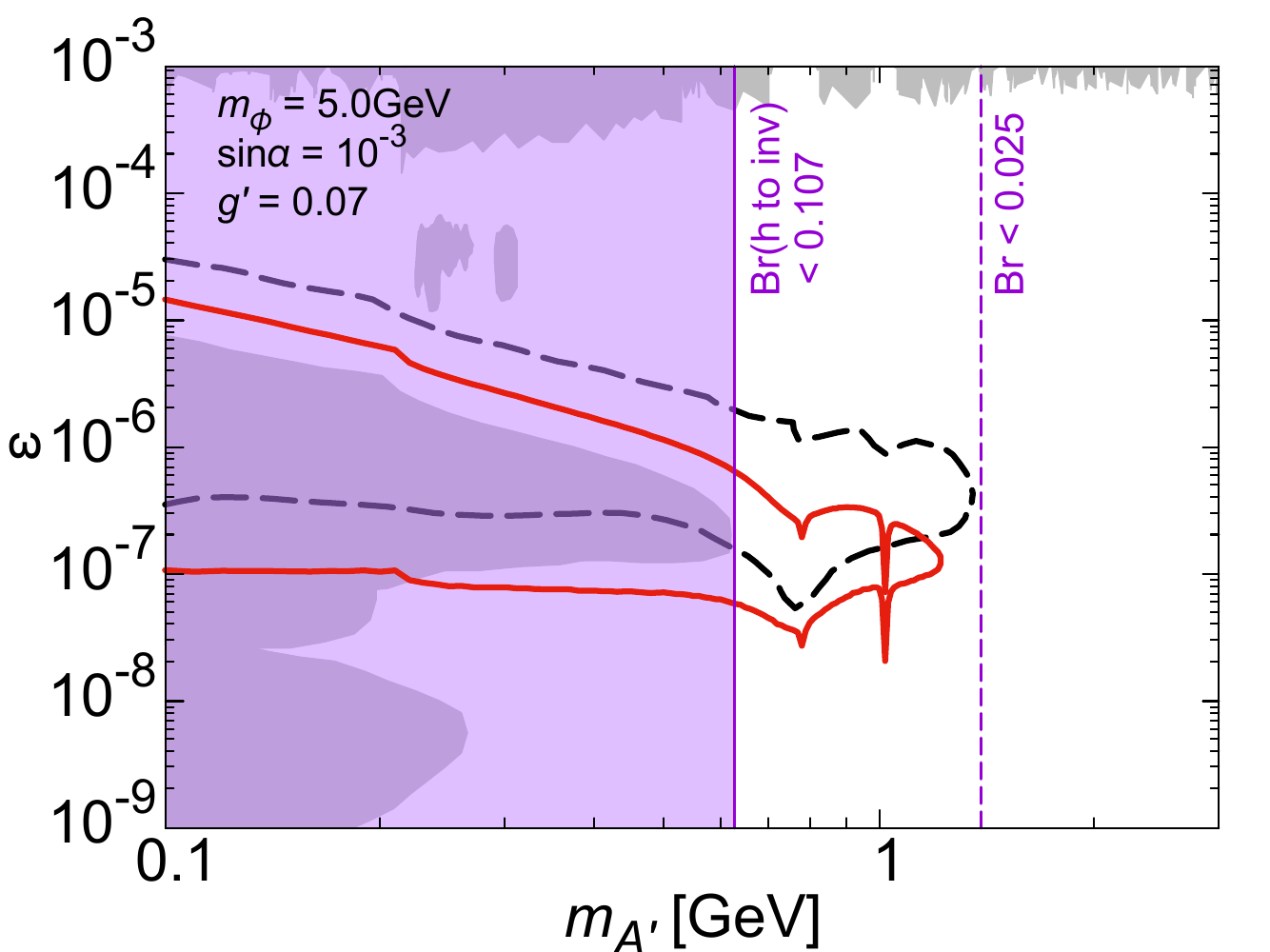}
\end{tabular}
    \caption{The $95$\% C.L. sensitivity contours at the FASER2 experiment for $\sin\alpha=10^{-3}$ and $m_\phi=5$ GeV. The gauge coupling is taken to $0.05$ (left) and $0.07$ (right), respectively. }
\label{fig:sensitivity3}
\end{figure}

After this paper was published, we realized that the scalar mixing can be taken up to $\mathcal{O}(10^{-3})$ in the case of invisible dark Higgs boson decay \cite{Ferber:2023iso}. This is the situation in our study that the dark Higgs boson dominantly decays into the pair of dark photons and the dark photon is long-lived.  Figure \ref{fig:sensitivity3} shows the FASER2 sensitivity region for $\sin\alpha=10^{-3}$ and $m_\phi=5$ GeV with $g'=0.05$ and $0.07$ , respectively. Taking the scalar mixing to $10^{-3}$, it is found that the FASER2 can explore the dark photon above $0.45$ GeV even for heavier dark Higgs boson than $B$ meson.

%
%======================================%
%<<<<<<<<<< ACKNOWLEDGMENTS >>>>>>>>>>>%
%======================================%

\section*{Acknowledgments}
We would like to thank Felix Kling for suggesting the possibility of large scalar mixing.
This work was partially supported by JSPS KAKENHI Grant Numbers JP24K07024 (T.~A.), JP23K13097 (K.~A.), JP22K03622 (T.~S.), JP23H01189 (T.~A. and T.~S.), and 
JST SPRING, Grant Number JPMJSP2136 (Y.~N.).

\appendix
\section{Decay Width of Dark Higgs Boson} \label{apdx:dark-higgs}
In this appendix, we give the decay widths of the dark Higgs boson into the SM fermions and dark photons.  
\begin{figure}[t]
\hspace{-10mm}
    \includegraphics[scale=0.27]{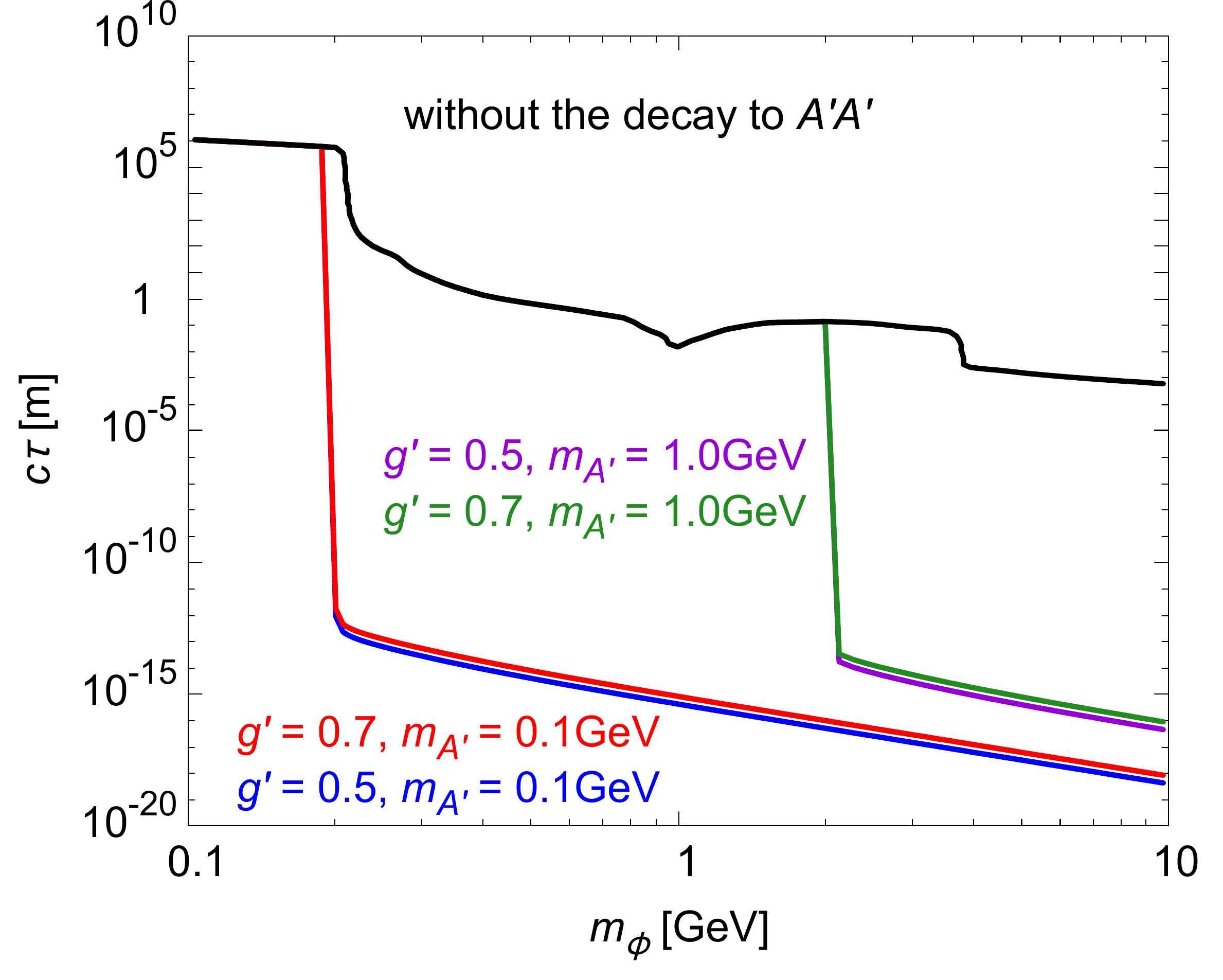}
    \caption{Decay length of the dark Higgs boson with respect to the dark Higgs mass. Red and blue curves correspond to $g' = 0.7$ and $0.5$ for $m_{A'} = 0.1$ GeV, while green and purple ones for $m_{A'} = 1.0$ GeV. Black curve represents the decay length only for the SM fermions and hadrons. The scalar mixing is fixed to $\alpha = 10^{-4}$.}
\label{fig:DecayLengthOfDarkHiggs}
\end{figure}

The decay widths are given by
\begin{align}
\label{eq:decaywidth-scalar-f}        
   \Gamma_{\phi \to f\bar{f}} &= \frac{m_\phi}{8 \pi} \left( \frac{m_f}{v} \right)^2 \sin^2\alpha 
        \left( 1 - \frac{4 m_f^2}{m_\phi^2} \right) \beta_\phi(f)~, \\
\label{eq:decaywidth-scalar-Ap}
   \Gamma_{\phi \to A'A'} &= \frac{{g'}^2 }{8 \pi} \frac{m_{A'}^2}{m_\phi} \beta_\phi(A') 
        \left( 2 + \frac{m_\phi^4}{4 m_{A'}^4} \left( 1 - \frac{2 m_{A'}^2}{m_\phi^2} \right)^2 \right)~,
\end{align}
where $m_\phi$ and $m_f$ are the mass of the dark Higgs and SM fermion, respectively, and $\beta_i(j) = \sqrt{1 - 4 m_j^2/m_i^2}$ is the kinematic factor of the decay $i \to jj$. 
The decay widths into the SM hadrons are complicated and found in refs.~\cite{Gunion:1989we,Donoghue:1990xh,McKeen:2008gd,Clarke:2013aya}. 
In our analysis, we used the data of the decay length for the dark Higgs decays into the SM particles implemented in FORESEE \cite{Kling:2021fwx}.

The decay length of the dark Higgs is shown in figure~\ref{fig:DecayLengthOfDarkHiggs} for $g'=0.7,~0.5$ and $m_{A'} = 0.1,~1.0$ GeV, respectively. One can see that the decay length becomes much shorter once the decay into the dark photon pair is open. 
The decay width of the dark Higgs into a pair of dark photons, eq.~\eqref{eq:decaywidth-scalar-Ap}, is significantly enhanced by the factor $m_\phi^2/m_{A'}^2$ when $m_\phi \gg m_{A'}$. 
On the contrary, the decay widths into the SM fermions, \eqref{eq:decaywidth-scalar-f}, are suppressed due to the scalar mixing $\sin^2\alpha$. Thus, the dark Higgs dominantly decays into a pair of dark photons once it is kinematically allowed. 
Furthermore, such a decay is prompt and the dark Higgs decays at close to the IP in the ATLAS detector.

%%%%%%%%%%%%%%%%%%%%%%%%%%%%%%%%%%%%%
\section{Decay Width of Dark Photon} \label{apdx:dark-photon}

\begin{figure}[t]
\hspace{-10mm}
    \includegraphics[scale=0.23]{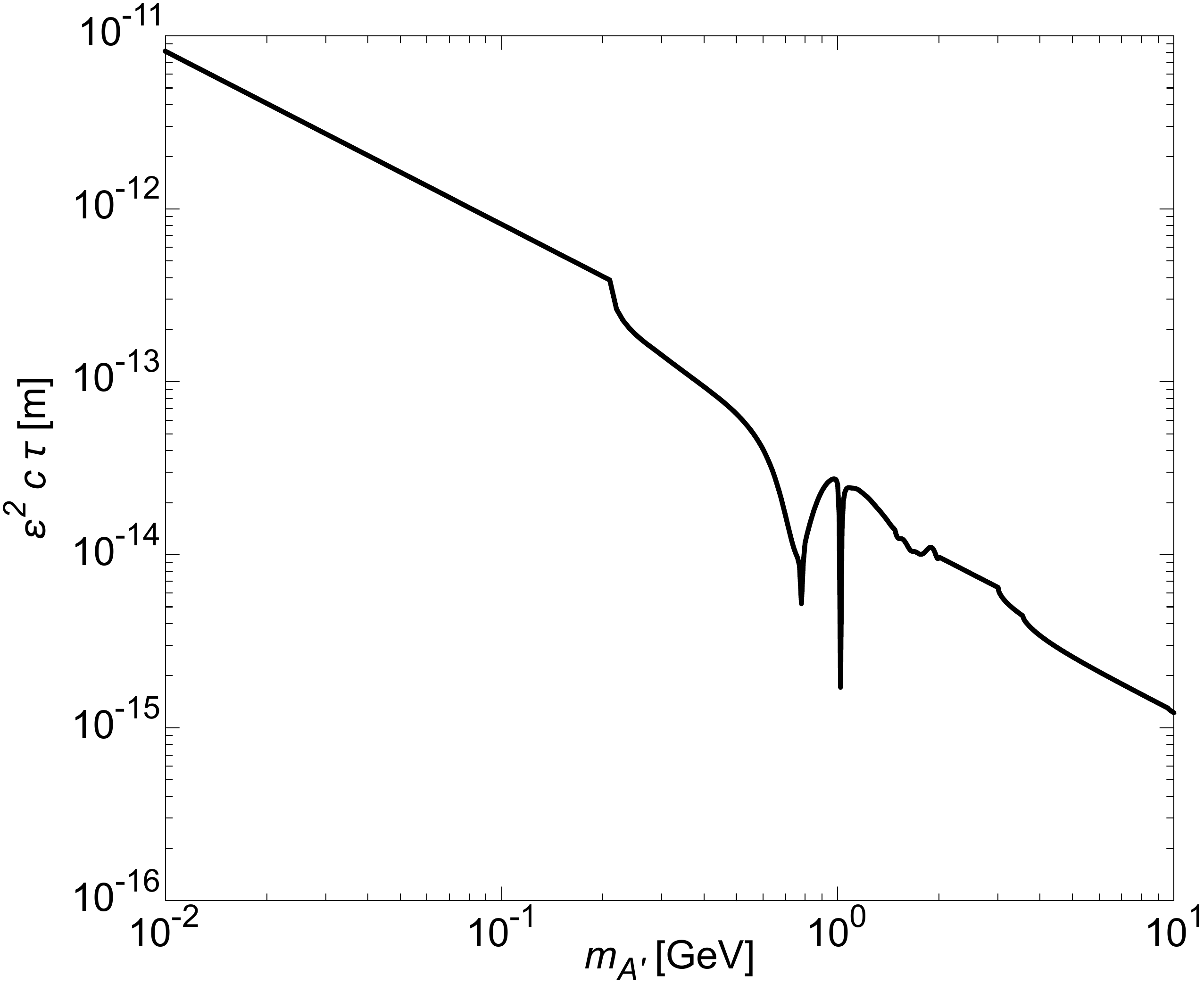}
    \caption{Decay length normalized by $\varepsilon^2$ of dark photon with respect to the dark photon mass.}
\label{fig:DecayLengthOfDarkPhoton}
\end{figure}
The decay width of the dark photon into the SM fermions can be calculated by using Eq.~\eqref{eq:dp-Jem},
\begin{align}
\label{eq:decaywidth-Ap-ff}
   \Gamma_{A' \to f\bar{f}} &= \frac{\varepsilon^2 e^2}{12 \pi} m_{A'} \left( 1 + 2 \frac{m_f^2}{m_{A'}^2} \right) \beta_{A'}(f)~, 
\end{align}
while the decay width into the SM hadrons is given by
\begin{align}
\label{eq:decaywidth-Ap-had}
   \Gamma_{A' \to {\rm hadrons}} &=\Gamma_{A' \to \mu \bar{\mu}} R(s=m_{A'}^2)~,
\end{align}
where $R(s) \equiv \sigma(e^+ e^- \to {\rm hadrons})/\sigma(e^+ e^- \to \mu \bar{\mu})$, being $s$ the center of mass energy, 
i.e. $m_{A'}$ for the decay. Similarly to the dark Higgs, we used the data of the 
decay length of the dark photon implemented in FORESEE \cite{Kling:2021fwx}.

The decay length of the dark photon is shown in figure~\ref{fig:DecayLengthOfDarkPhoton}. In the plot, decay length is normalized by $\varepsilon^2$. 
For instance, the decay length $c\tau$ is longer than about $1$ cm for $\varepsilon = 10^{-6}$ and $m_{A'} < 1$ GeV. The typical momentum of the dark photon is about $1$ TeV, and hence Lorentz boost factor is larger than $10^3$. Thus, the actual decay length is longer than $100$ m.

%%%%%%%%%%%%%%%%%%
%%% references %%%
%%%%%%%%%%%%%%%%%%
%\bibliographystyle{apsrev}
\bibliography{biblio}

\end{document}